\documentclass{article}
\usepackage[T1,T2A]{fontenc}
\usepackage[unicode]{hyperref}
\usepackage{amsmath,amssymb,amsthm,orcidlink,booktabs,pdflscape}
\usepackage{CJKutf8}
\usepackage[final]{microtype}
\title{Generalised wavefunction coefficients and acyclonesto-cosmohedra}
\author{Stefan Forcey\textsuperscript{\orcidlink{0000-0003-2251-5710}}\footnote{Department of Mathematics, University of Akron, Ohio, United States of America 44325}\and Ross Glew\textsuperscript{\orcidlink{0000-0002-8100-928X}}\footnote{Department of Physics, Astronomy and Mathematics, University of Hertfordshire, Hatfield, Hertfordshire\ \textsc{al10 9ab}, United Kingdom}\and Hyungrok Kim~(\begin{CJK*}{UTF8}{bsmi}金炯錄\end{CJK*})\textsuperscript{\orcidlink{0000-0001-7909-4510}}\footnotemark[2]
\\[1em]
\href{mailto:sforcey@uakron.edu}{\texttt{sforcey@uakron.edu}}
\hskip1em
\href{mailto:r.glew@herts.ac.uk}{\texttt{r.glew@herts.ac.uk}}
\\
\href{mailto:h.kim2@herts.ac.uk}{\texttt{h.kim2@herts.ac.uk}}
}
\usepackage[french,german,portuguese,italian,latin,russian,ukrainian,british]{babel}
\usepackage[osf]{newpxtext}
\usepackage[varbb,slantedGreek]{newpxmath}
\usepackage[artemisia]{textgreek}
\usepackage[utf8]{inputenc}
\usepackage{tikz-cd}
\usetikzlibrary{babel}
\hypersetup {
    pdftitle = {Generalised wavefunction coefficients and acyclonesto-cosmohedra},
    pdfauthor = {Stefan Forcey and Ross John Glew and Hyungrok Kim},
    pdfsubject = {hep-th, math.CO},
    pdflang = {en-GB},
    pdfkeywords = {acyclonestohedron, positive geometries, polytope, cosmohedron, oriented matroid},
}
\newcommand{\precdot}{\prec\mathrel{\mkern-5mu}\mathrel{\cdot}}

\begin{document}
\maketitle
\newtheorem{theorem}{Theorem}
\newtheorem{lemma}{Lemma}
\newtheorem{conjecture}{Conjecture}
\theoremstyle{definition}
\newtheorem{defn}{Definition}
\theoremstyle{remark}
\newtheorem{example}{Example}
\newcommand\cyrillic[1]{\fontfamily{Domitian-TOsF}\selectfont \foreignlanguage{russian}{#1}}
\begin{abstract}
Scattering amplitudes of  \(\operatorname{tr}(\phi^3)\) theory can be encoded as the canonical form of the Stasheff associahedron. Similarly, the flat-space wavefunction coefficients of the same theory are captured by the recently proposed cosmohedron, a non-simple polytope associated to the Stasheff associahedron; unitarity and locality of the amplitudes and wavefunction coefficients are then encoded in the factorisation properties of faces of these polytopes. In this paper, we argue that these desirable properties of the Stasheff associahedron are shared by a wider class of polytopes called acyclonestohedra and generalise the cosmohedron construction to arbitrary acyclonestohedra.

Acyclonestohedra are generalisations of Stasheff associahedra and graph associahedra defined on the data of a partially ordered set or, more generally, an acyclic realisable matroid on a building set. When the acyclonestohedron is associated to a partially ordered set, it may be interpreted as arising from Chan--Paton-like factors that are only (cyclically) partially ordered, rather than (cyclically) totally ordered as for the ordinary open string. In this paper, we argue that the canonical forms of acyclonestohedra encode scattering-amplitude-like objects that factorise onto themselves, thereby extending recent results for graph associahedra, and construct truncations of acyclonestohedra into acyclonesto-cosmohedra whose canonical forms may be interpreted as encoding a generalisation of the cosmological wavefunction coefficients.
As a byproduct, we provide evidence that acyclonesto-cosmohedra can be obtained as sections of graph cosmohedra. 
\end{abstract}
\tableofcontents
\section{Introduction and Summary}
Various functions that arise in physics --- scattering amplitudes, correlation functions and wavefunction coefficients --- are often characterised by factorisation channels. In certain kinematic limits, these functions become singular, and the residues at these singularities factorise into simpler components. This behaviour ultimately reflects the unitarity and locality of the underlying physical theory \cite{Arkani-Hamed:2016rak}. The positive geometries programme (reviewed in \cite{Ferro:2020ygk,Lam:2022yly}) aims to manifest these properties by encoding the physical quantity of interest as the canonical form of a positive geometry \cite{Arkani-Hamed:2017tmz,Arkani-Hamed:2013jha}. Within this framework, going to a factorisation channel corresponds to approaching a boundary facet of the positive geometry. We will focus on positive geometries given by convex polytopes. However, not every convex polytope is suitable in the sense that the faces may not factorise into smaller polytopes of the same class. Thus, one can ask the question: which classes of polytopes are `physics-like' in that their faces factorise as products of simpler polytopes of the same class?

In this article, we propose that \emph{acyclonestohedra} \cite{mpp,mppfull} provide a large class of such physics-like polytope and thus offer an approximate answer to this question. This family includes, as special cases, the classical associahedron which describes the scattering amplitudes of biadjoint \(\phi^3\) theory \cite{Arkani-Hamed:2017mur}, as well as the graph associahedra of \cite{devadoss2011deformations} that appear in  cosmological contexts \cite{Arkani-Hamed:2024jbp}. They also encompass the poset associahedra of \cite{galashin} which have not yet found application to physical processes.

\begin{figure}
    \centering
    \includegraphics[width=\linewidth]{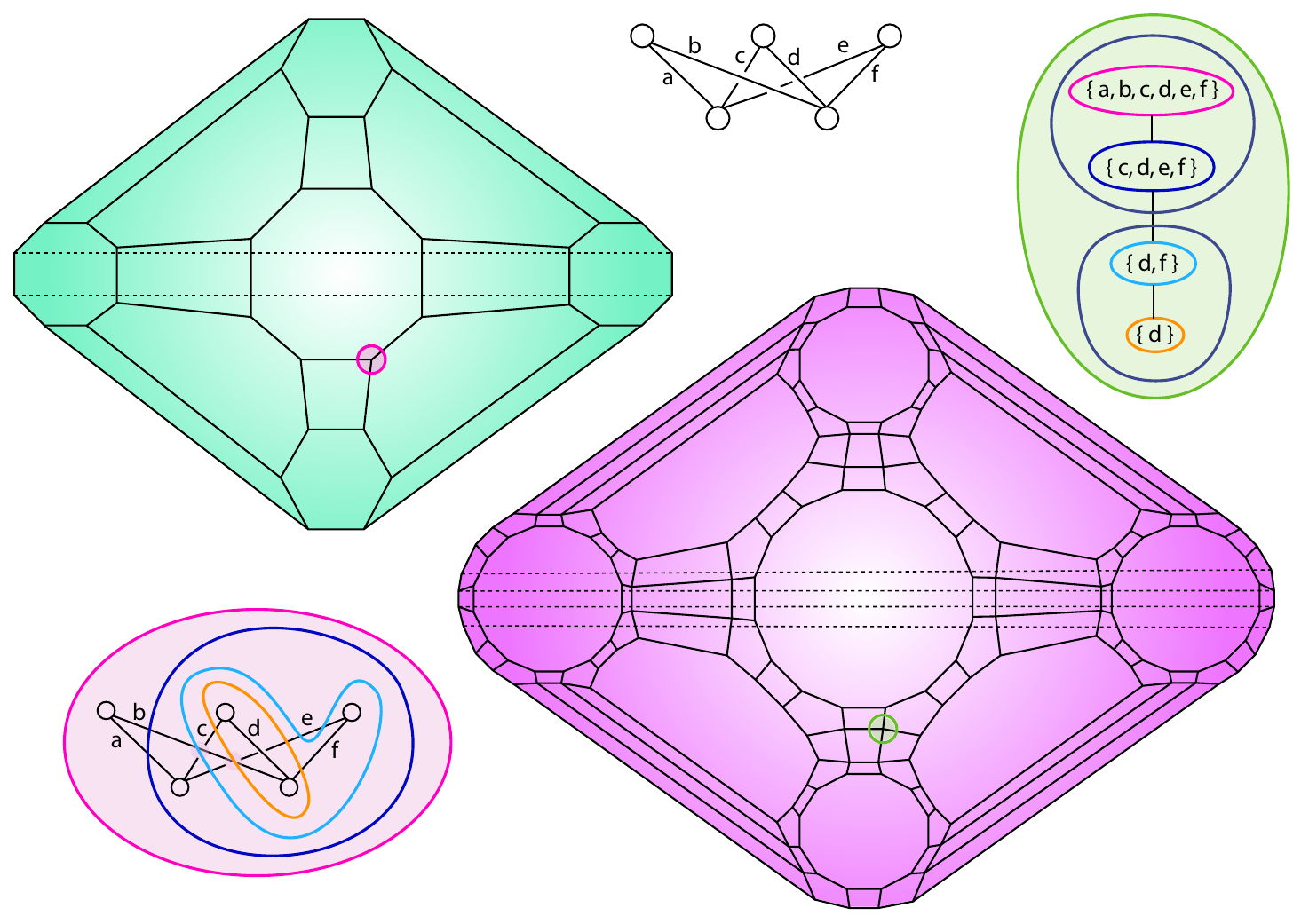}
    \caption{The acyclonestohedron (left) and associated acyclonesto-cosmohedron (right) for nestings of the poset $K_{2,3}$. The acyclonestohedron is shown as a realisation in \cite{sack}, and both polytopes as realisations here in Figure~\ref{fig:RealK23}. One vertex of each polytope is circled, with the corresponding  maximal nesting $\tau$ of $K_{2,3}$ shown below and maximal nested nesting \((\tau, \mathcal{N})\) 
    \label{fig:k23} shown above.}
\end{figure}

Furthermore, we generalise the construction of graph cosmohedra in \cite{Glew:2025otn} to define acyclonesto-cosmohedra; these further generalise the classical cosmohedron in \cite{Arkani-Hamed:2024jbp}. The key realisation being that for any polytope with faces indexed by nested sets, the nested sets themselves come equipped with a Hasse diagram which can be further imbued with nested sets, these ideas are advertised in Figure~\ref{fig:k23}. As a byproduct of our results, we provide evidence that acyclonesto-cosmohedra can be obtained as sections of graph cosmohedra, this generalises similar observations made for the acyclonestohedra \cite{mpp}.

This paper is organised as follows. In section~\ref{sec:physics_justification}, we argue that acyclonestohedra represent a general class of polytopal positive geometries manifesting locality and unitarity as well as other desirable physical properties. In section~\ref{sec:acyclonestohedra}, we review the definition of acyclonestohedra and associate rational functions, called \emph{amplitubes}, that reflect the exotic kinematics of the acyclonestohedra. In section~\ref{sec:cosmohedra}, we associate  generalisations of the cosmohedron to acyclonestohedra and present realisations and examples thereof.

\section{Polytopes that model physical processes}\label{sec:physics_justification}
As emphasised, in order for the canonical form of a convex polytope to model physical processes, there are several desired properties, all modelled on the Stasheff associahedron describing the scattering amplitudes of biadjoint cubic scalar field theory \cite{Arkani-Hamed:2017mur}:
\begin{itemize}
\item Its faces should factorise into products of smaller polytopes of the same class. For example, the faces of the Stasheff associahedron factorise into products of lower-dimensional associahedra.
\item It should be obtained as a compactification (or blow-up or truncation) of the interior of a simpler polytope modelling a colour or flavour structure. For example, the Stasheff associahedron is obtained as a compactification of the simplex \cite{Arkani-Hamed:2017mur}, which arises directly from the Chan--Paton factors of open strings.
\item It should have a construction in terms of intersections of half-spaces associated to Mandelstam-like variables and positive cut parameters. For example, the Stasheff associahedron can be obtained as an intersection of half-spaces associated to the planar Mandelstam variables \cite{Arkani-Hamed:2017mur}.
\end{itemize}
We argue that these criteria are all met by the class of acyclonestohedra.

\subsection{Locality and unitarity}
Factorisation is an essential ingredient of the positive-geometries programme, encompassing both unitarity and locality of physical systems. For graph associahedra \cite{cd}, it has been argued \cite{Glew:2025otn} that the locality is given by the fact that the denominators for the corresponding amplitude-like quantities (`amplitubes') correspond to connected subgraphs while unitarity corresponds to the fact that the residues of amplitubes are given by products of simpler amplitubes.

However, there exist classes of polytope beyond graph associahedra which also have this factorisation property. In one direction, the properties that characterise the tubes and tubings in terms of which graph associahedra are defined can be axiomatised into the notion of a building set; these define a generalisation of graph associahedra known as \emph{nestohedra} \cite{postnikov,fk,fs,zelevinsky}.\footnote{Nestohedra do \emph{not} generalise pseudograph associahedra \cite{cdf}, whose tubes do not form building sets.} In another direction, the \emph{poset associahedra} introduced in \cite{galashin} also have boundaries which factorise into lower poset associahedra, but in a way not captured by the axiomatics of building sets.
Subsequently, it was realised in \cite{mpp} that poset associahedra are naturally associated to building sets that carry additional orientation data, the authors axiomatised this structure into the notion of an \emph{oriented building sets}. The oriented building sets then yield a broader class of polytopes known as \emph{acyclonestohedra}, which encompass both the nestohedra and poset associahedra, see figure~\ref{fig:factorising_polytope_classes} and table~\ref{table:factorising_polytope_classes}. Thus, acyclonestohedra appear to us to be a large and natural class of polytopes that encapsulate the phenomenon of factorisation.

In fact, the definition of oriented building sets is easy to motivate, which we do so before launching into formal definitions in section~\ref{sec:acyclonestohedra}.
A poset may be represented by the corresponding Hasse diagram, which is a directed acyclic graph \(G\); the poset associahedron's faces correspond to a subset of the faces of the graph associahedron for the line graph \(L(G)\), where some would-be faces are eliminated according to a criterion depending on the orientation data of \(G\).
Thus, to generalise graph associahedra and nestohedra, one must keep track of the orientation data in a way that is consistent with factorisation --- that is, in some structure that behaves naturally according to restriction and contraction, the same way that building sets restrict and contract naturally (definition~\ref{def:building_set}). A combinatorialist will immediately realise that oriented matroids fit the bill, such that an oriented building set is a building set equipped with an oriented matroid (definition~\ref{def:oriented_building_set}).

Finally, we note in passing that every acyclonestohedron may be realised as a slice of a certain graph associahedron \cite[Def.~2.17]{mpp}, such that their amplitubes can always be realised as spurious poles inside certain representations of amplitubes for graph associahedra.

\begin{figure}\centering
\usetikzlibrary{backgrounds}
\begin{tikzpicture}
  \draw[thick] (-5,-3.5) rectangle (5,3.5);
  \node[anchor=north east] at (5,3.5) {acyclonestohedra};
  \begin{scope}
    \fill[blue!10] (-2,0) ellipse (2.5 and 2);
    \draw[blue, thick] (-2,0) ellipse (2.5 and 2);
    \node[blue] at (-3,2.3) {nestohedra};
  \end{scope}
  \begin{scope}
    \fill[green!20] (-1,0) ellipse (1.3 and 1.2);
    \draw[green!50!black, thick] (-1,0) ellipse (1.3 and 1.2);
    \node[green!50!black] at (-3,-0.7) {\shortstack{graph\\associahedra}};
  \end{scope}
  \begin{scope}[fill opacity=0.5]
    \fill[red!20] (1.8,0) ellipse (2.5 and 2);
    \draw[red!70!black, thick] (1.8,0) ellipse (2.5 and 2);
  \end{scope}
  \node[red!70!black] at (3,-2.3) {poset associahedra};
\end{tikzpicture}
\caption{Venn diagram of different classes of polytopes whose faces factorise. Acyclonestohedra include all other classes shown. Operahedra, as defined in \cite{laplante-anfossi}, are poset associahedra when the Hasse diagram of the poset is a tree. In that case tubings on the line graph are all compatible with the poset and they are thus included in the intersection of graph associahedra and poset associahedra.}\label{fig:factorising_polytope_classes}
\end{figure}

\begin{table}\centering
\begin{tabular}{rcc}
\toprule
& concrete (graph, poset) & abstract (set systems) \\\midrule
unoriented & graph associahedra \cite{cd} & \begin{tabular}{@{}c@{}}nestohedra\\ \cite{postnikov,fk,fs}\end{tabular}\\\addlinespace
partially oriented & \begin{tabular}{@{}c@{}} poset associahedra\\\cite{galashin}\end{tabular} & \begin{tabular}{@{}c@{}}acyclonestohedra\\ \cite{mpp}\end{tabular} \\\bottomrule
\end{tabular}

\caption{Comparison of different classes of polytopes whose faces factorise}\label{table:factorising_polytope_classes}
\end{table}

\subsection{Order polytope and generalised colour}
We wish to consider generalisations of the Stasheff associahedron that capture aspects of physics. The Stasheff associahedron can be obtained as a certain compactification or blowup of the interior of a simplex \cite{Arkani-Hamed:2017mur}, which captures aspects of colour structure.
We argue that a natural generalisation to this picture is provided by (generalised) \emph{order polytopes}, which are non-linear (in the sense of no longer being a linear order) generalisations of the simplex; their compactifications are the Galashin poset associahedra and, more generally, acyclonestohedra.

Let us first recall the origin of the simplex from colour symmetry. Suppose that we have a theory of massless fields that have a \(\mathfrak u(N)\) adjoint colour (or flavour) symmetry, so that we may work with colour-ordered \(n\)-point scattering amplitudes.
If the theory arises as a limit of an open-string theory in which the colour factor is realised by Chan--Paton factors at the boundary of the open string, then the tree-level amplitude is associated to the 
open string moduli space
\begin{equation}\mathcal M_{0,n}(\mathbb R)=
    \widetilde{\mathcal M}_{0,n}(\mathbb R)/\operatorname{SL}(2;\mathbb R),
\end{equation}
where \(\widetilde{\mathcal M}_{0,n}(\mathbb R)\) is the space of cyclic-order-preserving maps
\begin{equation}\label{eq:open-string-map}
    \sigma\colon\{1,\dotsc,n\}\to\mathbb{RP}^1,
\end{equation}
where \(\{1,\dotsc,n\}\) is given the standard cyclic order.
For \(n\ge3\),
the fundamental unit cell (amongst \((n-1)!/2\) unit cells \cite{devadoss}) in the \(\operatorname{SL}(2;\mathbb R)\) action may be identified by the gauge choice
\begin{align}
    \sigma\colon1&\mapsto0,&n-1&\mapsto1,&n&\mapsto\infty,
\end{align}
so that the unit cell is given by (ordinary) order-preserving maps
\begin{equation}
    \sigma\colon\{2,\dotsc,n-2\}\to(0,1),
\end{equation}
which is the simplex. The kinematic associahedron then naturally arises \cite{Arkani-Hamed:2017mur} as a natural compactification or blowup of (the interior of) the simplex.
That is, the simplex (and its compactification, the associahedron) arises as a natural consequence of the colour structure given by the open string.

This identification may also be seen directly from the bulk rather than the worldsheet. It is well known \cite{Kajiura:2001ng,Kajiura:2003ax,Doubek:2020rbg} that open string field theory (and quantum field theories obtained as its limits) are, in general, given by an \(A_\infty\)-algebra; more generally, in the homotopy-algebraic approach to scattering amplitudes \cite{Hohm:2017pnh,Jurco:2018sby,Jurco:2019bvp,Arvanitakis:2019ald,Jurco:2020yyu,Borsten:2021hua,Borsten:2022ouu}, the \(L_\infty\)-algebra of a quantum field theory theory with \(\mathfrak{su}(N)\) adjoint fields may be factorised into an \(A_\infty\)-algebra (the colour-stripped field theory) tensored with the colour Lie algebra \(\mathfrak{su}(N)\) \cite{Jurco:2019yfd,Borsten:2021hua}. The \(A_\infty\)-algebra, in turn, is governed by the \(A_\infty\)-operad, whose space of \((n-1)\)-ary operations (corresponding to \(n\)-point amplitudes) may be realised as the \((n-2)\)-dimensional Stasheff associahedron (see e.g.\ \cite{vallette}). Similarly, cyclohedra and permutohedra may be obtained as compactified moduli spaces of configurations \cite{ltv} and may be interpreted as corresponding to generalised colour.

We would like to extend this construction to a much broader class. An immediate choice is to generalise from the totally ordered set \(\{1,2,\dotsc,n\}\) in \eqref{eq:open-string-map} to an arbitrary cyclically partially ordered set (or cycloposet for short).\footnote{
A cyclic partial order on a set \(S\) is a ternary relation \(R(x,y,z)\) that is cyclically invariant (\(R(x,y,z)\iff R(y,z,x)\)), antisymmetric (if \(R(x,y,z)\), then not \(R(y,x,z)\)), and transitive (if \(R(x,y,w)\) and \(R(y,z,w)\), then \(R(x,z,w)\)) \cite{mueller,megiddo}.} Let \(\bar P\) be an arbitrary cycloposet, and consider the space of cyclic-order-preserving maps
\begin{equation}
    \sigma\colon\bar P\to\mathbb{RP}^1
\end{equation}
modulo \(\operatorname{SL}(2;\mathbb R)\). Now, using the special conformal generator of the \(\operatorname{SL}(2;\mathbb R)\) gauge symmetry, we may pick a particular element \(\top\in P\) and map \(\sigma(\top)=\infty\). Then \(P\coloneqq \bar P\setminus\{\top\}\) forms a poset (with partial order \(i\prec j\) iff \(R(i,j,\top)\)), and \(\sigma\) reduces to an order-preserving map
\begin{equation}
    \sigma\colon P\to\mathbb R.
\end{equation}
Using the translation generator of \(\operatorname{SL}(2;\mathbb R)\), we may require that
\begin{equation}
    \sum_{i\in P}\sigma(i)=0.
\end{equation}
Finally, using the dilatation generator of \(\operatorname{SL}(2;\mathbb R)\), we may require that (assuming \(\preceq\) is nontrivial)
\begin{equation}
    \sum_{\substack{i,j\in P\\i\precdot j}}\sigma(j)-\sigma(i)=1,
\end{equation}
where \(i\precdot j\) (read as `\(i\) is covered by \(j\)') means that \(i\prec j\) and that there does not exist a \(k\in P\) such that \(i\prec k\prec j\) \cite[p.~13]{aigner} \cite[p.~11, §1.14]{dp}. The space of such maps then forms the \emph{order polytope} \cite{stanley,galashin}, which is a convex simple polytope of dimension \(\#P-2\).\footnote{More precisely, this gives Galashin's generalisation \cite{galashin} of the order polytope; when \(P\) is bounded above and below, this reduces to Stanley's original definition \cite{stanley} (see \cite[Rem.~3.1]{sack}).} When \(\bar P\) is a totally cyclically ordered set, this reduces to the simplex found in open strings. A natural compactification / blow-up of the interior of the order polytope then produces Galashin's poset associahedra \cite{galashin},\footnote{These are not to be confused with the earlier Devadoss--Forcey--Reisdorf--Showers poset associahedra \cite{dfrs}; it is not clear whether the faces of the latter class factorise due to the filledness condition.} which form a large class of acyclonestohedra.

We can also directly interpret the `generalised colour' given by order polytopes directly from a bulk field-theoretic picture rather than from the open string worldsheet. For a field theory with only adjoint fields \(\phi_1,\phi_2,\dotsc\), one writes the Lagrangian as, for example,
\begin{equation}
    \mathcal L = \int \operatorname{Tr}\left(\partial_\mu\phi_i\partial^\mu\phi^i+\lambda_{ijk}\phi_i\phi_j\phi_k+\lambda_{ijkl}\phi_i\phi_j\phi_k\phi_l+\dotsb\right).
\end{equation}
The entire action is enclosed inside a big trace, and inside it one may treat the adjoint-valued fields \(\phi_i\) as non-commuting variables; modulo cyclicity of the trace, the Lagrangian is a noncommutative associative polynomial of the fields (with coefficients in differential operators). When gauge symmetry (and thus the Batalin--Vilkovisky formalism) is involved, one should relax the associativity up to homotopy; the different ways of multiplying terms, such as
\begin{equation}
    \left((\phi_i\phi_j)\phi_k\right)\phi_l\sim (\phi_i\phi_j)(\phi_k\phi_l)\sim \phi_i\left(\phi_j(\phi_k\phi_l)\right)\sim\dotsb,
\end{equation}
then define the vertices of the Stasheff associahedron, and the homotopies that relate them define the faces of the Stasheff associahedron. Thus, one may only multiply symbols that are adjacent --- that is, that correspond to edges in a path graph. One may then replace the path graph with a more general graph; then one may interpret the colour structure as ways to multiply generalised words, where one may multiply symbols only along edges \cite[§2.3]{sack} (cf.\ \cite[§2.1]{laplante-anfossi}).

Poset associahedra, however, do not encompass examples, such as the cyclohedron, which also arise as compactifications of moduli spaces \cite{ltv}.
More generally, as generalisations of simplices and order polytopes, we may consider positive cells (topes) in a hyperplane arrangement.
Suppose that we have a collection of vectors \(v_1,\dotsc,v_k\in V=\mathbb R^n\). These vectors then induce a collection of hyperplanes in the dual space \(V^*\), where each vector \(v_i\) corresponds to the hyperplane
\begin{equation}
    H_i\coloneqq \{\lambda\in V^*\,|\,\lambda(v_i)=0\}\subset V^*.
\end{equation}
Furthermore, each hyperplane comes with a \emph{sign}, that is, \(V^*\) is partitioned into \(V^*=H_i^-\cup H_i\cup H_i^+\) where
\begin{equation}
    H_i^\pm\coloneqq\{\lambda\in V^*\,|\,\pm\lambda(v_i)>0\}.
\end{equation}
The data of such `signed' hyperplanes defines a \emph{realisable oriented matroid} on the set \(\{v_1,\dotsc,v_k\}\); this oriented matroid is \emph{acyclic} if the \emph{positive tope} \(\bigcap_iH_i^+\subseteq V^*\) is nonempty.
The positive tope then serves as a generalisation of the simplex and the order polytope: the order polytope is the special case where the realisable oriented matroid arises as the oriented matroid associated to the Hasse digraph of a poset.
The acyclonestohedron is then obtained as a compactification or blowup of this positive tope \cite[Thm.~4.1]{mpp}.

\subsection{Kinematic variables}
In a Poincaré-invariant field theory in \(d\) spacetime dimensions, for an \(n\)-point scattering amplitude with incoming momenta \(p_1,\dotsc,p_n\), one can form the \(\binom n2\) Mandelstam variables \(s_{ij}=p_i\cdot p_j\).
If \(d\ge n-1\), then the complete set of identities amongst them consists of the \(n\) conditions of the form
\begin{equation}\sum_{j\in\{1,\dotsc,n\}\setminus\{i\}}p_i\cdot p_j=0\end{equation}
for each \(i\).
Thus, there are \(\binom n2-n=\frac12n(n-3)\) independent Mandelstam variables.
If the particles all transform under the adjoint representation of a colour symmetry, then one can require that \(p_1,\dotsc,p_n\) be cyclically colour-ordered, and then these independent variables are naturally parameterised by the \(\frac12n(n-3)\) planar Mandelstam variables
\begin{equation}
    X_{i,j}\coloneqq (p_i+\dotsb+p_{j-1})^2\qquad(1\le i<j\le n,\;i+1\ne j,\;(i,j)\ne(1,n)).
\end{equation}
The fact that these planar Mandelstam variables are manifestly positive means that the kinematic associahedron (for biadjoint \(\phi^3\) theory) is obtained by a truncation of the simplex
\begin{equation}
    X_{i,j}\ge0.
\end{equation}
These planar Mandelstam variables are then in bijection with the tubes of the path graph \(L(P_{n-1})=P_{n-2}\)), thus providing a natural kinematic interpretation of the \((n-3)\)-dimensional Stasheff associahedron; these tubes are certain connected intervals in the totally ordered set with \(n-1\) elements, whose Hasse diagram is \(P_{n-1}\).

When we generalise from the path graph \(P_{n-1}\) (the Hasse diagram of the totally ordered set with \(n-1\) elements, or the cyclically totally ordered set with \(n\) elements with an element at `infinity' removed) to Hasse diagrams of arbitrary posets (corresponding to Galashin poset associahedra), we continue to have tubes associated to certain connected intervals\footnote{namely, sets \(T\) such that whenever \(i,j\in T\) and \(i\prec k\prec j\), then \(k\in T\); these are called \emph{convex subsets} in the poset literature}; these may be considered poset generalisations of the planar Mandelstam variables.
In the full generality of arbitrary acyclonestohedra, this condition is abstracted to even more in terms of oriented matroids and building sets; however, one may still interpret a \(k\)-dimensional acyclonestohedron as describing a \(k+3\)-point process (the \(3\) corresponds to the dimension of \(\operatorname{SL}(2;\mathbb R)\) that can be gauge-fixed); the dimension \(k\) is given by the dimension of the span of the vectors realising the oriented matroid.

Note that, in general, the number of independent kinematic variables (i.e.\ number of facets) for an \((n-3)\)-dimensional acyclonestohedron can be greater than or less than \(\frac12n(n-3)\); for example, at \(n=5\) points (that is, in two dimensions), both the triangle (i.e.\ simplex) and hexagon (i.e.\ permutahedron) are possible in addition to the pentagon (i.e.\ Stasheff associahedron). 

\section{Acyclonestohedra and their realisation}\label{sec:acyclonestohedra}
Having argued that acyclonestohedra are a natural class of physics-like geometries,
we now move on to the formal definition of these polytopes. As we will show the canonical forms of the acyclonestohedra allow us to define amplitude-like functions called \emph{amplitubes}. We illustrate these ideas through multiple examples.

\subsection{Building sets and nestohedra}
We begin by defining building sets and nestings, which axiomatise the notion of tubes and tubings on a graph. In general, the terms nesting, nested set, tubing, and piping are closely related: the first two are synonymous and the the second two are specializations to graphs and posets respectively.

\begin{defn}[\cite{postnikov,fk,fs}]\label{def:building_set}
    A \emph{building set} \(\mathcal B\) on a ground set \(S\) is a collection of nonempty subsets of \(S\) such that
    \begin{itemize}
        \item for any \(s\in S\), then \(\{s\}\in\mathcal B\);
        \item whenever \(B,B'\in\mathcal B\) with \(B\cap B'\ne\varnothing\), then \(B\cup B'\in\mathcal B\).
    \end{itemize}
    A \emph{connected component} of a building set \((S,\mathcal B)\) is an inclusion-maximal element of \(\mathcal B\); the set of connected components is denoted by \(\max(\mathcal B)\subseteq\mathcal B\).
    A \emph{nesting} \(\mathcal N\) of a building set \(\mathcal B\) is a subset \(\max(\mathcal B)\subseteq\mathcal N\subseteq\mathcal B\) such that
    \begin{itemize}
        \item whenever \(B,B'\in\mathcal N\), then either \(B\subseteq B'\) \(B'\subseteq B\) or \(B\cap B'=\varnothing\);
        \item for any finite collection of pairwise disjoint elements \(B_1,\dotsc,B_k\in\mathcal N\) (with \(k>1\)), then \(B_1\cup\dotsb\cup B_k\not\in\mathcal B\).
    \end{itemize}
    The collection of nestings $\{ \mathcal{N}\setminus\max(\mathcal{B})\,|\, \mathcal{N} \text{ is nesting} \}$ under reverse inclusion define the poset of faces of a convex polytope called the \emph{nestohedron}. The facets of the nestohedron are labelled by nestings of the form \(\{B\}\) for \(B\in\mathcal B\). These facets factorise into products of two nestohedra defined on the {\it restriction} and {\it contraction} of $\mathcal{B}$ to $\{ B \}$. Where, for any subset \(R\subseteq S\), the \emph{restriction} \(\mathcal B_{\,|R}\) and \emph{contraction} \(\mathcal B_{/R}\) of $\mathcal{B}$ to $S$ are defined as the building sets
    \begin{align}
        \mathcal B_{\,|R}&\coloneqq\{B\in\mathcal B\,|\,B\subseteq R\},&
        \mathcal B_{/R}&\coloneqq\{B\setminus R\,|\,R\not\supseteq B\in\mathcal B\}.
    \end{align}
\end{defn}

\begin{example}[\cite{cd}]
    Let \(G\) be a simple graph with vertex set \(S=\operatorname V(G)\). We define the graphical building set \(\mathcal B\) to be the collection of subsets of \(S\) whose induced subgraphs of \(G\) are connected. The corresponding nestohedron, as defined in \cite{postnikov}, is the \emph{graph associahedron} of \(G\).
    The connected components of the graphical building set correspond to the connected components of the graph.\footnote{Note that the original definition in \cite{cd} allowed proper subsets of connected components to be included in each nesting. Under the present definition this is equivalent to including in the building set the base set \(S\), so that \(S\) itself becomes the only connected component. Thus, for instance, we find the simplex from a single edge hypergraph; see Example~\ref{ex:simplex}.}
    Given a set of vertices \(R\subseteq S\), then \(\mathcal B_{\,|R}\) and \(\mathcal B_{/R}\) are the graphical building sets of the induced subgraph of \(R\) in \(G\) and the reconnected complement of \(R\) in \(G\).
\end{example}

\subsection{Oriented matroids}
A matroid can be seen as an abstraction of both linear independence in vector spaces and acyclic edge subsets in graphs, while oriented matroid can be seen as a refinement that incorporates direction, generalising vectors over ordered fields and acyclic edge subsets in directed graphs \cite{blvswz}. Note, oriented matroids and their connection to amplitudes have recently been studied in \cite{Lam:2024jly,2025arXiv250220782E}.

\begin{defn}\label{def:oriented_matroid}
    A \emph{signed set} \(X=(X,\sigma\) is a \(\mathbb Z_2\)-graded set, i.e.\ a set \(X\) together with a an assignment of signs \(\sigma\colon X\to\{+1,-1\}\) to every element. We may formally write such a set as \(X=X^+-X^-=x_1+x_2+\dotsb-y_1-y_2-\dotsb\) where \(x_1,x_2,\dotsc\in X^+\) are the elements with degree \(+1\) and \(y_1,y_2,\dotsc\in X^-\) are the elements with degree \(-1\); thus \(-X=X^--X^+\) is the signed set with all degrees reversed.
    An \emph{oriented matroid} \((S,\mathcal C)\) on a finite set \(S\) is a collection of signed sets (called \emph{signed circuits}) \(\mathcal C\) such that
    \begin{itemize}
        \item \(\varnothing\not\in\mathcal C\)
        \item if \(C\in\mathcal C\), then \(-C\in\mathcal C\)
        \item if \(X\in\mathcal C\ni Y\), and \(X^+\cup X^-=Y^+\cup Y^-\), then \(X=Y\) or \(X=-Y\)
        \item if \(X,Y\in\mathcal C\) with \(X\ne-Y\) and \(s\in X^+\cap Y^-\), then there exists a \(Z\in\mathcal C\) such that \(Z^\pm\subset(X^\pm\cup Y^\pm)\setminus\{e\}\).
    \end{itemize}
    Given a subset \(R\subseteq S\), the \emph{restriction} \((S,\mathcal C)_{\,|R}\) and \emph{contraction} \((S,\mathcal C)_{/R}\) are the oriented matroids given by
    \begin{align}
        (S,\mathcal C)_{\,|R}&\coloneqq (R,\{C\in\mathcal C\,|\,C^+\cup C^-\subseteq R\}),\\
        (S,\mathcal C)_{/R}&\coloneqq \left(S\setminus R,\left\{(C^+\setminus R)-(C^-\setminus R)\,|\,C\in\mathcal C\right\}\right)        
    \end{align}
    respectively.
    An oriented matroid is \emph{acyclic} if it does not have a signed circuit whose elements are all positive.
\end{defn}

\begin{example}[Oriented matroid from matrix]
    Let \(M\) be an \(m\times n\) matrix over \(\mathbb R\) (or, more generally, any ordered field). Consider the collection of linear dependencies of the columns of \(M\), i.e.\ coefficients \(\lambda\in\mathbb R^n\) such that \(M\lambda=0\); some of them are minimal in that \(\lambda\) has the fewest nonzero components under the obvious partial ordering. Let \(S=\{1,\dotsc,n\}\). For each minimal linear dependency, we can associate the signed set whose positive elements are \(i\in S\) such that \(\lambda_i>0\) and whose negative entries are \(j\in S\) such that \(\lambda_j<0\). The collection of signed sets associated to minimal linear dependencies forms an oriented matroid over \(S\).
    An oriented matroid definable from a matrix is called \emph{realisable}.
\end{example}

\begin{example}[Oriented matroid from hyperplane arrangement]
    The preceding example can be more geometrically phrased in terms of a hyperplane arrangement as follows.
    Let \(V\) be a finite-dimensional real vector space, and let \(a_1,\dotsc,a_k\in V^*\) be a finite set of linear functionals on \(V\). This defines a hyperplane arrangement \(\{\ker a_1,\dotsc,\ker a_k\}\) on \(V\) together with the choice of a positive half-space \(H_i^+=\{v\in V\,|\,a_i(v)>0\}\) for each hyperplane \(H_i=\ker a_i\).
    The collection of hyperplanes partitions \(V\) into strata depending on whether the values of \((a_1,\dotsc,a_k)\) are positive, negative, or zero, which correspond to the \emph{covectors} of the associated oriented matroid; circuits may be defined in terms of covectors \cite{blvswz}.
\end{example}

\begin{example}[Oriented matroid from directed graph]
    Let \(G\) be a directed pseudograph (i.e.\ self-loops and multiple parallel arcs are allowed). Then the \emph{graphical oriented matroid} is the matroid on the set of arcs \(\operatorname E(G)\) whose circuits \(C\) are minimal circuits, with \(C_+\) being the edges oriented along the circuit and \(C_-\) being the edges oriented opposite to the circuit. Restriction and contraction correspond to restriction and contraction of a subset of arcs.
    The graphical oriented matroid is acyclic if and only if the directed pseudograph is acyclic.
    Every oriented matroid arising from a directed pseudograph is realisable.
\end{example}

\subsection{Oriented building sets and acyclonestohedra}
We now finally define oriented building sets and acyclonestohedra, their ABHY-like realisations, and their associated amplitubes.

\begin{defn}[\cite{mpp}]\label{def:oriented_building_set}
    An \emph{oriented building set} \((S,\mathcal B,\mathcal C)\) is a building set \((S,\mathcal B)\) together with an oriented matroid \((S,\mathcal C)\) on the same ground set \(S\). An \emph{acyclic nesting} of an oriented building set \((S,\mathcal B,\mathcal C)\) is a nesting \(\mathcal N\subset\mathcal B\) of \((S,\mathcal B)\) such that, for every \(B\in\mathcal N\), the oriented matroid \(\left((S,\mathcal C)_{\,|B}\right)_{/\bigcup\{N\in\mathcal N\,|\,N\subsetneq B\}}\)
    is acyclic (the notation \(\bigcup\) means the union of all elements of a collection of sets).
    When \((S,\mathcal C)\) is realisable,
    the collection \(\{\mathcal N\setminus\max\mathcal B\,|\,\mathcal N\text{ is an acyclic nesting}\}\) under reverse inclusion is the poset of faces of a convex polytope called the \emph{acyclonestohedron} of \((S,\mathcal B,\mathcal C)\).\footnote{The fact that this poset actually corresponds to a convex polytope is not obvious, but is shown in \cite{mpp} under the assumption that \(\mathcal C\) is realisable.}
\end{defn}

From the definition, it follows that the unique codimension $0$ face (the interior of the polytope) is the unique nesting \(\max\mathcal B\) (which is trivially acyclic), whilst the facets (codimension $1$ faces) are in canonical bijection with those sets \(B\in\mathcal B\) such that the oriented matroids \((S,\mathcal C)_{\,|B}\) and \((S,\mathcal C)_{/B}\) are both acyclic, and the vertices (maximal-codimension faces) are in canonical bijection with acyclic nestings that are maximal under inclusion.

If the oriented matroid \((S,\mathcal C)\) is realised by the vectors \((a_i)_{i\in S}\) that span a \(k\)-dimensional vector space, the dimension of the acyclonestohedron is given by \(k - |\max\mathcal B|\), where \(|\max\mathcal B|\) is the number of connected components of \(\mathcal B\).
Given a realisable oriented building set \((S,\mathcal B,\mathcal C)\) and a facet given by \(B\in\mathcal B\) such that \((S,\mathcal C)_{\,|B}\) and \((S,\mathcal C)_{/B}\) are both acyclic, then the facet of the acyclonestohedron corresponding to \(B\) factorises as
\begin{multline*}
    \text{facet for \(B\)} = \text{acyclonestohedron for \((S,\mathcal B_{\,|B},\mathcal C_{\,|B})\)} \times \\
    \text{acyclonestohedron for \((S,\mathcal B_{/B},\mathcal C_{/B})\)}.
\end{multline*}
This may be applied recursively to higher-codimension faces.

\begin{example}
    Suppose that \((S,\mathcal B)\) is a building set (for example, given by a graph or hypergraph). Then one can always take the trivial oriented matroid \(\mathcal C=\varnothing\), which can be realised by an \(S\)-indexed collection of linearly independent vectors. This trivialises the acyclicity condition so that the acyclonestohedron reduces to the nestohedron for \((S,\mathcal B)\); the graph associahedron \cite{cd} is a special case.
\end{example}

\begin{example}[\cite{galashin}]
    Given a poset \(P\), let the set of its covers be \(S\coloneqq\{(i,j)\in P^2\,|\,i\precdot j\}\); this is, equivalently, the set of edges of the Hasse diagram \(G\) of \(P\). On \(S\), we may construct the building set \((S,\mathcal B)\) associated to the line graph \(L(G)\) as well as the realisable oriented matroid \((S,\mathcal C)\) associated to the digraph structure of \(G\). Then the acyclonestohedron corresponding to the oriented building set \((S,\mathcal B,\mathcal C)\) is the Galashin poset associahedron for \(P\).
\end{example}

\subsection{ABHY-like realisations of acyclonestohedra and amplitubes}
The mathematics literature \cite{sack,mpp,mppfull} contains realisations of acyclonestohedra in terms of intersections of half-spaces that generalise the ABHY-like realisations of graph associahedra given in \cite{Glew:2025otn}, in terms of which we may define amplitude-like functions, called \emph{amplitubes}, defined from the associated canonical forms.

Suppose that we are given an oriented building set \((S,\mathcal B,\mathcal C)\) and that \(\mathcal C\) is realised by a collection of vectors \(a_i\in V^*\) that span a finite-dimensional real vector space \(V^*\). For each \(B\in\mathcal B\) such that \((S,\mathcal C)_{\,|B}\) and \((S,\mathcal C)_{/B}\) are both acyclic, define the kinematic variable \(X_B\colon V\to\mathbb R\) as the affine function
\begin{equation}\label{eq:acyclonestohedron_function}
    X_B = \sum_{i\in B}a_i - \sum_{\substack{B'\in\mathcal B\\B'\subseteq B}}c_{B'},
\end{equation}
where the \(c_B\) are nonnegative real numbers (the \emph{cut parameters}) for each \(B\in\mathcal B\), chosen such that \(c_B>0\) is a positive real number whenever \(B\) contains more than one element and \(c_B=0\) whenever \(B\) contains only one element. In particular, we have a cut parameter \(c_\kappa\) for each connected component \(\kappa\in\max\mathcal B\).
Note that the \(X_B\) are not linearly independent if the vectors \(a_i\) realising the oriented matroid \((S,\mathcal C)\) are not linearly independent; one has the relations
\begin{equation}\label{eq:linear_dependence_amongst_X}
    \sum_{i\in B}\lambda_i\left(X_B-\sum_{\substack{B'\in\mathcal B\\B'\subseteq B}}c_{B'}\right)=0\text{ whenever }\sum_{i\in B}\lambda_ia_i=0.
\end{equation}
Then the ABHY-like realisation of the acyclonestohedron is given by the set of points \(v\in V\) such that
\begin{equation}\label{eq:acyclonestohedron_realisation}
\begin{aligned}
    X_B(v)&\ge0\text{ for every \(B\in\mathcal B\) such that \((S,\mathcal C)_{\,|B}\) and \((S,\mathcal C)_{/B}\) are acyclic}\\
    X_\kappa(v)&=0\text{ for every \(\kappa\in\max\mathcal B\)}.
\end{aligned}
\end{equation}
This manifestly generalises the ABHY-like realisation for graph associahedra given in \cite{Glew:2025otn}. On the other hand, when the vectors realising the oriented matroid are not all linearly independent, we must impose additional conditions on the cut parameters \(c_B\) in addition to their positivity --- that is, in physical terms, there are additional exotic constraints amongst the Mandelstam variables. A sufficient condition to satisfy these exotic kinematic constraints is to impose
\begin{equation}\label{eq:sufficient_condition_acyclonestohedron}
    c_B \ll c_{B'}
\end{equation}
whenever \(|B\,|<\,|B'|\); to be precise, it suffices to have \(c_{B'}/c_B\le R\), where \(R>1\) is a certain constant depending only on \((S,\mathcal B,\mathcal C)\) \cite[Def.~2.16]{mpp}.

Given this realisation, we may define the \emph{amplitube} associated to the realisable oriented building set \((S,\mathcal B,\mathcal C)\) as
\begin{equation}\label{eq:amplitube}
    A_{(S,\mathcal B,\mathcal C)} = \sum_\tau\prod_{B\in\tau}\frac1{X_B},
\end{equation}
where the sum ranges over the acyclic nestings of \((S,\mathcal B,\mathcal C)\) that are maximal under inclusion (i.e.\ the vertices of the acyclonestohedron). This manifestly generalises the amplitube of the graph associahedron given in \cite{Glew:2025otn}.

Just like the amplitubes for graph associahedra, locality and unitarity are manifest for amplitubes for general acyclonestohedra in that denominators correspond to connected subsets according to the poset topology \cite{wachs} of \(\mathcal B\) (or, for Galashin poset associahedra, subgraphs of the edge graph of the Hasse diagram) and that residues factorise.
Note that the acyclonestohedron amplitubes often \emph{lack} certain poles that would have existed in the corresponding nestohedron but are ruled out by the acyclicity condition. 

Since the acyclonestohedron is a simple polytope, the amplitube \eqref{eq:amplitube} may be recovered from the canonical form \(\Omega\) of the acyclonestohedron as
\begin{equation}
    \Omega = A_{(S,\mathcal B,\mathcal C)}\,\bigwedge_B\mathrm dX_{\{B\}},
\end{equation}
where the wedge product ranges over all singleton acyclic nestings except for \(|\max\mathcal B|\) many singleton acyclic nestings, which we may eliminate by solving the equality constraints in \eqref{eq:acyclonestohedron_realisation}.

\subsection{Examples of acyclonestohedra and their realisations}
Let us consider some examples. Examples~\ref{ex:stasheff}, \ref{ex:simplex}, and \ref{ex:permutohedron} simply reproduce nestohedra (these examples are also found in \cite{Glew:2025otn}).
On the other hand, Examples~\ref{ex:diamond} and \ref{ex:bowtie} showcase the more general acyclonestohedra.

\begin{example}[Stasheff associahedron]\label{ex:stasheff}
    Let \(P\) be a totally ordered set consisting of \(n+1\) elements and \(n\) covers \(\mathtt e_1,\dotsc,\mathtt e_n\):
    \begin{equation}
        P = \begin{tikzcd}[column sep=0,row sep=1em]
            \bullet\dar[dash,"\mathtt e_n"]\\\vdots\dar[dash,"\mathtt e_3"]\\\bullet\dar[dash,"\mathtt e_2"]\\\bullet\dar[dash,"\mathtt e_1"]\\\bullet
        \end{tikzcd}.
    \end{equation}
    That is, the Hasse diagram is the path graph \(P_{n+1}\).
    The ground set is the set of covers:
    \begin{equation}
        S = \{\mathtt e_1,\dotsc,\mathtt e_n\}.
    \end{equation}
    The acyclic oriented matroid contains no signed circuits (\(\mathcal C=\varnothing\)) and can be realised by any set of \(n\) linearly independent vectors.
    The building set is the graphical building set associated to the line graph \(L(P_{n+1})=P_n\) of the Hasse diagram of \(P\):
    \begin{equation}
        \mathcal B
        =\left\{
            \{\mathtt e_i,\dotsc,\mathtt e_{j-1}\}
        \,|\,
            1\le i<j\le n+1
        \right\}.
    \end{equation}
    Since the oriented matroid is trivial, the corresponding acyclonestohedron is the same as the graph associahedron of the path graph \(P_n\), which is the \((n-1)\)-dimensional Stasheff associahedron that describes \((n+2)\)-point scattering amplitudes of the biadjoint scalar field. For example, for \(n=3\), the building set is
    \begin{equation}
        \mathcal B = \left\{\{\mathtt e_1\},\;\{\mathtt e_2\},\;\{\mathtt e_3\},\{\mathtt e_1,\mathtt e_2\},\{\mathtt e_2,\mathtt e_3\},\{\mathtt e_1,\mathtt e_2,\mathtt e_3\}\right\}.
    \end{equation}
    The ABHY-like realisation of the acyclonestohedron is given by
    \begin{gather}
        X_{\{\mathtt e_1\}}\ge0,\quad
        X_{\{\mathtt e_2\}}\ge0,\quad
        X_{\{\mathtt e_3\}}\ge0,\notag\\
        X_{\{\mathtt e_1\}}+X_{\{\mathtt e_2\}}\ge c_{\{\mathtt e_1,\mathtt e_2\}},\quad
        X_{\{\mathtt e_2\}}+X_{\{\mathtt e_3\}}\ge c_{\{\mathtt e_2,\mathtt e_3\}}\\
        X_{\{\mathtt e_1\}}+X_{\{\mathtt e_2\}}+X_{\{\mathtt e_3\}}=c_{\{\mathtt e_1,\mathtt e_2\}}+c_{\{\mathtt e_2,\mathtt e_3\}}+c_{\{\mathtt e_1,\mathtt e_2,\mathtt e_3\}}.\notag
    \end{gather}
    This is therefore the two-dimensional associahedron, which is a pentagon.
    When the cut parameters \(c_{12},c_{23}\to0\), this reduces to the order polytope, namely the two-dimensional simplex (i.e.\ triangle).
    The corresponding five-point amplitube is
    \begin{equation}
        \mathcal A =
        \frac1{X_{\{\mathtt e_1\}}X_{\{\mathtt e_1,\mathtt e_2\}}}
        +\frac1{X_{\{\mathtt e_2\}}X_{\{\mathtt e_1,\mathtt e_2\}}}
        +\frac1{X_{\{\mathtt e_2\}}X_{\{\mathtt e_2,\mathtt e_3\}}}
        +\frac1{X_{\{\mathtt e_3\}}X_{\{\mathtt e_2,\mathtt e_3\}}}
        +\frac1{X_{\{\mathtt e_1\}}X_{\{\mathtt e_3\}}},
    \end{equation}
    which can be identified as the tree-level five-point double-partial amplitude of the biadjoint scalar theory when the two colour orderings agree. The example for $n=4$ which is the three-dimensional associahedron is pictured in Figure~\ref{fig:stasheff_cosmohedron}.
\end{example}

\begin{example}[Simplex]\label{ex:simplex}
Consider the building set associated to the trivial connected hypergraph on \(n\) vertices \(S=\{\mathtt v_1,\dotsc,\mathtt v_1\}\), all contained within a single hyperedge (and no other hyperedges).
The corresponding building set is
\begin{equation}
    \mathcal B = \left\{\{\mathtt v_1\},\dotsc,\{\mathtt v_n\},S\right\}
\end{equation}
on the ground set \(S\).
Consider the trivial oriented matroid on \(S\) with no signed circuits, which may be realised by any set of \(n\) linearly independent vectors.
Since the oriented matroid is trivial, the corresponding acyclonestohedron is simply the nestohedron of \(\mathcal B\), which is the simplex \(\triangle^n\).
The ABHY-like realisation is given by the standard simplex
\begin{equation}X_{\{\mathtt e_1\}}\ge0,\quad\dotsc,\quad X_{\{\mathtt e_n\}}\ge0,\quad X_{\{\mathtt e_1\}}+\dotsb+X_{\{\mathtt e_n\}}=c_S.\end{equation}
The corresponding \((n+2)\)-point amplitube is given by
\begin{equation}
    \mathcal A_n
    =\frac{X_{\{\mathtt e_1\}}+\dotsb+X_{\{\mathtt e_n\}}}{X_{\{\mathtt e_1\}}X_{\{\mathtt e_2\}}\dotsm X_{\{\mathtt e_n\}}}.
\end{equation}
\end{example}

\begin{example}[Permutohedron]\label{ex:permutohedron}
    Consider the claw poset of \(n+1\) elements and \(n\) covers:
    \begin{equation}P_n=
    \begin{tikzcd}[]
    \bullet\ar[rrd,dash,"\mathtt e_1"']&\bullet\drar[dash,"\mathtt e_2"]&\dotsm\dar[dash]&\bullet\dlar[dash]&\bullet\ar[lld,dash,"\mathtt e_n"]\\
    &&\bullet
    \end{tikzcd}.\end{equation}
    The ground set is the set of covers
    \begin{equation}
        S = \{\mathtt e_1,\dotsc,\mathtt e_n\}.
    \end{equation}
    The building set is that associated to the complete graph on \(S\) (i.e.\ the line graph of the Hasse diagram of \(P_n\)), consisting of all nonempty (possibly improper) subsets of \(S\):
    \begin{equation}
        \mathcal B = \left\{B\subseteq S\,|\,B\ne\varnothing\right\}.
    \end{equation}
    The corresponding oriented matroid has no signed circuits, i.e.\ it may be realised by any set of \(n\) linearly independent vectors. Since the oriented matroid is trivial, the corresponding acyclonestohedron is simply the nestohedron of \(\mathcal B\) (i.e.\ graph associahedron of the complete graph \(K_n\)), which is the permutohedron.
The corresponding \((n+2)\)-point amplitube is given by
\begin{equation}
    \mathcal A_n=\sum_{\sigma\in S_n}
    \frac1{X_{\{\mathtt e_{\sigma(1)}\}}X_{\{\mathtt e_{\sigma(1)},\mathtt e_{\sigma(2)}\}}\dotsm X_{\{\mathtt e_{\sigma(1)},\mathtt e_{\sigma(n)}\}}},
\end{equation}
where \(S_n\) is the set of all permutations of \(\{1,\dotsc,n\}\).
For example, at \(n=3\), the ABHY-like realisation is given by
    \begin{equation}
    \begin{gathered}
        X_{\{\mathtt e_1\}}\ge0,\;X_{\{\mathtt e_2\}}\ge0,\;X_{\{\mathtt e_3\}}\ge0,\\
        X_{\{\mathtt e_1\}}+X_{\{\mathtt e_2\}}\ge c_{\{\mathtt e_1,\mathtt e_2\}},\;
        X_{\{\mathtt e_2\}}+X_{\{\mathtt e_3\}}\ge c_{\{\mathtt e_2\mathtt e_3\}},\;
        X_{\{\mathtt e_1\}}+X_{\{\mathtt e_3\}}\ge c_{\{\mathtt e_1,\mathtt e_3\}},\\
        X_{\{\mathtt e_1\}}+X_{\{\mathtt e_2\}}+X_{\{\mathtt e_3\}}=c_{\{\mathtt e_1,\mathtt e_2\}}+c_{\{\mathtt e_2,\mathtt e_3\}}+c_{\{\mathtt e_1,\mathtt e_3\}}+c_{\{\mathtt e_1,\mathtt e_2,\mathtt e_3\}},
    \end{gathered}
    \end{equation}
    and we have the amplitube
\begin{multline}
    \mathcal A_3
    =\frac1{X_{\{\mathtt e_1\}}X_{12}}+\frac1{X_{\{\mathtt e_2\}}X_{\{\mathtt e_1,\mathtt e_2}\}}
    +\frac1{X_{\{\mathtt e_1\}}X_{\{\mathtt e_1,\mathtt e_3\}}}\\
    +\frac1{X_{\{\mathtt e_3\}}X_{\{\mathtt e_2,\mathtt e_3\}}}
    +\frac1{X_{\{\mathtt e_2\}}X_{\{\mathtt e_2,\mathtt e_3\}}}+\frac1{X_{\{\mathtt e_3\}}X_{\{\mathtt e_2,\mathtt e_3\}}}.
\end{multline}
\end{example}

\begin{example}[Hexagon from diamond poset]\label{ex:diamond}
    Consider the diamond poset
    \begin{equation}P_\diamond=
    \begin{tikzcd}[column sep=0,row sep=1em]
    &\bullet\dlar[dash,"\mathtt a"']\drar[dash,"\mathtt b"]\\\bullet\drar[dash,"\mathtt c"']&&\bullet\dlar[dash,"\mathtt d"]\\
    &\bullet
    \end{tikzcd}.\end{equation}
    The corresponding ground set of the oriented building set is the set of covers:
    \begin{equation}
        S = \{\mathtt a,\mathtt b,\mathtt c,\mathtt d\}.
    \end{equation}
    The oriented matroid on the ground set \(S\) is given by the two signed circuits:
    \begin{equation}
        \mathcal C =  \pm \left\{
            \mathtt a
            +\mathtt c -\mathtt d -\mathtt b
        \right\}.
    \end{equation}
    It is realised by vectors \(\mathbf v_{\mathtt a}\), \(\mathbf v_{\mathtt b}\), \(\mathbf v_{\mathtt c}\), and \(\mathbf v_{\mathtt d}\) such that \(\mathbf v_{\mathtt a}+\mathbf v_{\mathtt c}=\mathbf v_{\mathtt b}+\mathbf v_{\mathtt d}\) but where every subset of three vectors is linearly independent.
    The building set on \(S\) is given by the collection of all nonempty subsets of \(S\) except for two disconnected ones:
    \begin{equation}
        \mathcal B = \mathcal P(S) \setminus \left\{\varnothing,\{\mathtt a,\mathtt d\},\{\mathtt b,\mathtt c\}\right\},
    \end{equation}
    where \(\mathcal P(S)\) denotes the power set of \(S\).
    There are 13 nestings:
    \begin{itemize}
        \item one trivial nesting: \(\{S\}\)
        \item six facets, of the form \(\{S,B\}\) for \(B\in\mathcal B\setminus \left\{\{\mathtt a,\mathtt c\},\{\mathtt b,\mathtt d\}\right\}\) with \(|B|\le2\). The would-be facets \(\{S,\{\mathtt a,\mathtt c\}\}\) and \(\{S,\{\mathtt b,\mathtt d\}\}\) violate the acyclicity condition.
        \item six vertices:
        \[\begin{gathered}
        \{S,\{\mathtt a,\mathtt b\},\{\mathtt a\}\}\,\qquad\{S,\{\mathtt a,\mathtt b\},\{\mathtt b\}\},\qquad\{S,\{\mathtt c,\mathtt d\},\{\mathtt c\}\},\\\{S,\{\mathtt c,\mathtt d\},\{\mathtt d\}\},\qquad\{S,\{\mathtt a\},\{\mathtt d\}\},\qquad\{S,\{\mathtt b\},\{\mathtt c\}\}.
        \end{gathered}\]
    \end{itemize}
    The corresponding acyclonestohedron is thus a hexagon (see \cite[Fig.~3b]{mpp}).
    The ABHY-like realisation is as follows:
    \begin{equation}
    \begin{gathered}
        X_{\{\mathtt a\}}\ge 0,\qquad X_{\{\mathtt b\}}\ge 0,\qquad X_{\{\mathtt c\}}\ge0,\qquad X_{\{\mathtt d\}}\ge0,\\
        X_{\{\mathtt a\}}+X_{\{\mathtt b\}}\ge c_{\{\mathtt a,\mathtt b\}},\qquad
        X_{\{\mathtt c\}}+X_{\{\mathtt d\}}\ge c_{\{\mathtt c,\mathtt d\}},\\
        X_{\{\mathtt a\}}+X_{\{\mathtt b\}}+X_{\{\mathtt c\}}+X_{\{\mathtt d\}}=c_{\{\mathtt a,\mathtt b\}}+c_{\{\mathtt c,\mathtt d\}}
        +c_{\{\mathtt a,\mathtt b,\mathtt c,\mathtt d\}},\\
        X_{\{\mathtt a\}}+X_{\{\mathtt c\}}-X_{\{\mathtt b\}}-X_{\{\mathtt d\}}=0.
    \end{gathered}
    \end{equation}
    This simplifies to the square with two corners truncated:
    \begin{equation}
    \begin{gathered}
        0\le X_{\{\mathtt a\}}\le C,\qquad
        0\le X_{\{\mathtt b\}}\le C\\
        X_{\{\mathtt a\}}+X_{\{\mathtt b\}}\ge c_{\{\mathtt a,\mathtt b\}},\qquad
        (C-X_{\{\mathtt a\}})+(C-X_{\{\mathtt b\}})\ge c_{\{\mathtt c,\mathtt d\}},
    \end{gathered}
    \end{equation}
    where
    \begin{equation}\label{eq:square_size_for_diamond}
        C\coloneq \frac12\left(c_{\{\mathtt a,\mathtt b\}}+c_{\{\mathtt c,\mathtt d\}}
        +c_{\{\mathtt a,\mathtt b,\mathtt c,\mathtt d\}}\right),
    \end{equation}
    and \(X_{\{\mathtt c\}}=C-X_{\{\mathtt a\}}\) and \(X_{\{\mathtt d\}}=C-X_{\{\mathtt b\}}\).
    For this to be a valid realisation, the cut parameters must satisfy the kinematic constraint
    \begin{equation}\label{eq:kinematic_constraint_diamond}
        \left\lvert c_{\{\mathtt a,\mathtt b\}}-c_{\{\mathtt c,\mathtt d\}}\right\rvert<c_{\{\mathtt a,\mathtt b,\mathtt c,\mathtt d\}}.
    \end{equation}
    The amplitube associated to the diamond poset is then
    \begin{equation}
        \mathcal A
        =
        \frac1{X_{\{\mathtt a\}}X_{\{\mathtt{a,b}\}}}
        +
        \frac1{X_{\{\mathtt b\}}X_{\{\mathtt{a,b}\}}}
        +\frac1{X_{\{\mathtt c\}}X_{\{\mathtt{c,d}\}}}
        +
        \frac1{X_{\{\mathtt d\}}X_{\{\mathtt{c,d}\}}}
        +\frac1{X_{\{\mathtt a\}}X_{\{\mathtt d\}}}
        +\frac1{X_{\{\mathtt b\}}X_{\{\mathtt c\}}}.
    \end{equation}
\end{example}

\begin{example}[Octagon from bowtie poset]\label{ex:bowtie}
    Consider the bowtie poset
    \begin{equation}
        P_{\bowtie} = \begin{tikzcd}
            \bullet\dar[dash,"\mathtt a"']\drar[dash,"\mathtt d" very near start]&\bullet\dar[dash,"\mathtt c"]\dlar[dash,"\mathtt b" very near end]\\
            \bullet&\bullet
        \end{tikzcd}.
    \end{equation}
    The ground set, the set of signed circuits, and the building set are respectively
    \begin{align}
        S &= \{\mathtt{a,b,c,d}\}&
        \mathcal C&=\{
        \mathtt{\pm a\mp b\pm c\mp d}
        \}&
        \mathcal B&=\mathcal P(S)\setminus
        \left\{
            \{\mathtt{a,c}\},
            \{\mathtt{b,d}\},
            \varnothing
        \right\},
    \end{align}
    where \(\mathcal P(S)\) is the power set of \(S\).
    There are 17 nestings:
    \begin{itemize}
    \item one trivial nesting: \(\{S\}\),
    \item eight facets: \(\{S,T\}\) for \(T\in\mathcal B\) with \(1\le|T|\le2\),
    \item eight vertices: \(\{S,T,T'\}\) for \(T\in\mathcal B\) with \(|T|=2\) and \(T'\subset T\) with \(|T'|=1\).
    \end{itemize}
    Thus the corresponding acyclonestohedron is an octagon, as pictured in Figure~\ref{fig:bowtie}. The ABHY-like realisation is
    \begin{equation}
    \begin{gathered}
        X_{\{\mathtt a\}}\ge0,\quad X_{\{\mathtt b\}}\ge0,\quad X_{\{\mathtt c\}}\ge0,\quad X_{\{\mathtt d\}}\ge0,\\ X_{\{\mathtt a\}}-X_{\{\mathtt b\}}+X_{\{\mathtt c\}}-X_{\{\mathtt d\}}=0,\\
        X_{\{\mathtt a\}}+X_{\{\mathtt b\}}\ge c_{\{\mathtt a,\mathtt b\}},\quad
        X_{\{\mathtt b\}}+X_{\{\mathtt c\}}\ge c_{\{\mathtt b,\mathtt c\}},\\
        X_{\{\mathtt c\}}+X_{\{\mathtt d\}}\ge c_{\{\mathtt c,\mathtt d\}},\quad
        X_{\{\mathtt a\}}+X_{\{\mathtt d\}}\ge c_{\{\mathtt d,\mathtt a\}},\\
        X_{\{\mathtt a\}}+X_{\{\mathtt b\}}+X_{\{\mathtt c\}}+X_{\{\mathtt d\}}=c_{\{\mathtt{a,b}\}}+c_{\{\mathtt{b,c}\}}+c_{\{\mathtt{c,d}\}}+c_{\{\mathtt{d,a}\}}+c_{\{\mathtt{a,b,c,d}\}}.
    \end{gathered}
    \end{equation}
    This simplifies to the truncated square
    \begin{equation}
    \begin{gathered}
        (X_{\{\mathtt a\}},X_{\{\mathtt b\}})\in [0,C]\times[0,C],\\
        X_{\{\mathtt a\}}+X_{\{\mathtt b\}}\ge c_{\{\mathtt a,\mathtt b\}},\;
        (C-X_{\{\mathtt a\}})+X_{\{\mathtt b\}}\ge c_{\{\mathtt c,\mathtt b\}},\\
        X_{\{\mathtt a\}}+(C-X_{\{\mathtt b\}})\ge c_{\{\mathtt a,\mathtt d\}},\;
        (C-X_{\{\mathtt a\}})+(C-X_{\{\mathtt b\}})\ge c_{\{\mathtt c,\mathtt d\}},
    \end{gathered}
    \end{equation}
    where \(C\coloneq \frac12(c_{\{\mathtt{a,b}\}}+c_{\{\mathtt{b,c}\}}+c_{\{\mathtt{c,d}\}}+c_{\{\mathtt{d,a}\}}+c_{\{\mathtt{a,b,c,d}\}})\).
    This correctly realises the acyclonestohedron (octagon) as long as the cut parameters obey the constraint
    \begin{equation}\label{eq:kinematic_constraint_bowtie}
        \max\left\{\left\lvert c_{\{\mathtt{a,b}\}}+c_{\{\mathtt{b,c}\}}-c_{\{\mathtt{c,d}\}}-c_{\{\mathtt{d,a}\}}\right\rvert,\left\lvert c_{\{\mathtt{b,c}\}}+c_{\{\mathtt{c,d}\}}-c_{\{\mathtt{d,a}\}}-c_{\{\mathtt{a,b}\}}\right\rvert\right\}<c_{\{\mathtt{a,b,c,d}\}}.
    \end{equation}
    The corresponding amplitube is
    \begin{multline}
        \mathcal A
        =\frac1{X_{\{\mathtt a\}}X_{\{\mathtt a,\mathtt b\}}}
        +
        \frac1{X_{\{\mathtt b\}}X_{\{\mathtt a,\mathtt b\}}}
        +
        \frac1{X_{\{\mathtt b\}}X_{\{\mathtt b,\mathtt c\}}}
        +
        \frac1{X_{\{\mathtt c\}}X_{\{\mathtt b,\mathtt c\}}}\\
        +
        \frac1{X_{\{\mathtt c\}}X_{\{\mathtt c,\mathtt d\}}}
        +
        \frac1{X_{\{\mathtt d\}}X_{\{\mathtt c,\mathtt d\}}}
        +
        \frac1{X_{\{\mathtt d\}}X_{\{\mathtt d,\mathtt a\}}}
        +
        \frac1{X_{\{\mathtt a\}}X_{\{\mathtt d,\mathtt a\}}}.
    \end{multline}
\end{example}

\section{Acyclonesto-cosmohedra}\label{sec:cosmohedra}
In this section, we associate to every acyclonestohedron a non-simple polytope called the \emph{acyclonesto-cosmohedron} that generalises the cosmohedron for Stasheff associahedra \cite{Arkani-Hamed:2024jbp} and graph cosmohedra \cite{Glew:2025otn} as well as rational functions called \emph{cosmological amplitubes} that generalise the cosmological amplitubes for graph associahedra \cite{Glew:2025otn}. The acyclonesto-cosmohedra are also called poset cosmohedra in the poset case.
For the Stasheff case, amplitubes are known to be closely related to wavefunction coefficients for Bunch--Davies vacua in flat space as well as Friedman--Lemaître--Robertson--Walker metrics \cite{Arkani-Hamed:2024jbp}, and we may expect that the acyclonesto-cosmohedra are related to wavefunction coefficients in cosmological states with exotic kinematics whose scattering amplitudes are described by the amplitubes for the acyclonestohedra.

\subsection{Definition of acyclonesto-cosmohedra}\label{ssec:cosmohedron_definition}
Intuitively, in a cosmohedron, each face of the original positive geometry is refined into a poset of faces. Since faces in the acyclonestohedron correspond to nestings, it follows that we are to associate a nesting to a nesting, that is, to construct nested nestings; the poset of such nested nestings then define the acyclonesto-cosmohedron.

More concretely, recall that, for any acyclic nesting \(\tau\subseteq\mathcal B\) on an oriented building set \((S,\mathcal B,\mathcal C)\), the elements of \(\tau\) are partially ordered by inclusion. The Hasse diagram of the poset \((\tau,\subseteq)\) is a rooted forest due to the requirement of elements in \(\tau\) to be pairwise nested or disjoint, with the roots given by \(\max\mathcal B\). Since we are dealing with forests (acyclic graphs), the orientation does not matter, and we may simply consider the building set on the line graph \(L(G_\tau)\) of the Hasse diagram \(G_\tau\) of \((\tau,\subseteq)\). This naturally leads to the following definition.

\begin{defn}
    Given a building set \((S,\mathcal B)\),
    a \emph{nested nesting} \((\tau,\mathcal N)\) is a nesting \(\tau\subseteq\mathcal B\) together with a nesting \(\mathcal N\subseteq\mathcal P(\{(i,j)\in\tau\times\tau\,|\,i\precdot j\})\) on (the Hasse diagram of) the poset \((\tau,\subseteq)\).
Nested nestings are ordered by operations of collapsing a nest that is minimal in the poset \(\mathcal N\) (the edges are contracted and the nodes are identified, and given the label of the largest nest) or discarding a non-maximal nest of \(\mathcal N\).
That is, given two nested nestings \((\tau,\mathcal N)\) and \((\tau',\mathcal N')\), then \((\tau',\mathcal N')\preceq(\tau,\mathcal N)\) means that \(\mathcal N'\) is formed from \(\mathcal N\) by repeatedly collapsing a minimal nest or discarding a non-maximal nest. Note that this implies that \(\tau'\subseteq\tau\). 
    The acyclonesto-cosmohedron for the realisable oriented building set \((S,\mathcal B,\mathcal C)\) is a polytope whose poset of faces is equivalent to the poset of nested nestings on \((S,\mathcal B,\mathcal C)\) (with the relation \(\preceq\) reversed). 
\end{defn}

An acyclic nesting \(\tau\) of the acyclonestohedron may be identified with the nested nesting \((\tau,\operatorname{conn}(L(G_\tau)))\), where \(G_\tau\) is the Hasse diagram of \((\tau,\subseteq)\) and \(\operatorname{conn}(L(G_\tau))\) is the (collection of sets of vertices of) connected components of the line graph of \(G_\tau\) (or, equivalently, the collection of sets of edges of each connected components of \(G_\tau\), ignoring one-vertex connected components).

\begin{figure}
    \centering
    \includegraphics[width=\linewidth]{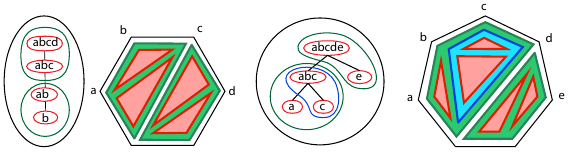}
    \caption{The maximal nested nestings here are on the building set from a path graph or poset as described in Example~\ref{ex:stasheff}. They are paired with their corresponding pictures of subpolygon collections from \cite{Arkani-Hamed:2024jbp}.}
    \label{fig:biject}
\end{figure}

This is superficially different from the definition based on `regions' in previous literature \cite{Arkani-Hamed:2024jbp,Glew:2025otn}; however, explicit computation shows that they agree. The regions associated to a nested nesting \((\tau,\mathcal N)\) are in bijection with the elements of \(\mathcal N\); each \(N\in\mathcal N\) is a set consisting of pairs \((i,j)\in\tau\times\tau\) with \(i\preceq j\), and the region corresponding to \(N\) is then the `union' of the formal differences \(j\setminus i\).
For the case of the classical cosmohedron of \cite{Arkani-Hamed:2024jbp}, we show the correspondence between collections of subpolygons (Russian dolls) and nested nestings in Figure~\ref{fig:biject}.

An advantage of the present definition is that it generalises readily: one can consider nested nested nestings, nested\textsuperscript3 nestings, and so on, to obtain \emph{iterated cosmohedra} (if such iterated nestings in fact are polytopal).

The acyclonesto-cosmohedron also satisfies a factorisation property generalising that given in \cite{Glew:2025otn}. Each facet of the acyclonesto-cosmohedron \(C_{(S,\mathcal B,\mathcal C)}\) for the oriented nested complex \((S,\mathcal B,\mathcal C)\) corresponds (via the identification just mentioned) to a nesting \(\tau\subseteq\mathcal B\), and we have the factorisation for the corresponding facet \(\mathcal{F}_\tau\):
 \begin{Large} \begin{equation}
      \mathcal{F}_\tau
    = A_\tau
    \times \prod_{B\in\tau}C_{((S,\mathcal B,\mathcal C)_{|B})_{/\bigcup\{N\in\tau\,|\,N\subsetneq B\}}}
\end{equation}\end{Large} 
where \(A_\tau\) is the poset associahedron for the poset \((\tau,\subseteq)\)\footnote{Or, equivalently, the graph associahedron for the line graph of the Hasse diagram of \((\tau,\subseteq)\); this line graph is called the \emph{spine} in \cite[(3.10)]{Glew:2025otn}.} and \(C_{((S,\mathcal B,\mathcal C)_{\,|B})_{/\bigcup\{N\in\mathcal \tau\,|\,N\subsetneq B\}}}\) is the acyclonesto-cosmohedron associated to the oriented building set given by
\(((S,\mathcal B,\mathcal C)_{\,|B})_{/\bigcup\{N\in\tau\,|\,N\subsetneq B\}}\) (this is the same restriction--contraction found in definition~\ref{def:oriented_building_set}).

\subsubsection{Face combinatorics and simplicity}
 In what follows we will assume that the building set has only one connected component. For any nested nesting $(\tau,\mathcal{N})$ the codimension of the face of the acyclonesto-cosmohedron labelled by that nested nesting is the number of nests in $\mathcal{N},$ always including the improper nest. That is, the dimension of a face is $d-\,|\mathcal{N}|$.  
Then
any facet \((\tau,\mathcal{N})\) of an acyclonesto-cosmohedron has \(\mathcal{N} = \{\tau\}\). We often simply draw this as the nesting \(\tau\). A facet which corresponds to a maximal nesting $\tau$ is combinatorially equivalent to the Galashin poset associahedron $A_{\tau}$. A poset of nests in a nesting always has a Hasse diagram which is a tree. Thus as shown in \cite{defant2024operahedronlattices} a facet equivalent to $A_{\tau}$ is in fact an \textit{operahedron}, as defined in \cite{laplante-anfossi}. In fact, all the poset associahedra $A_{\tau}$ occurring in the factorisation of faces just described are operahedra.

We use the same facts to bound the degree of any vertex of the acyclonesto-cosmohedron and to generalise the fact mentioned in \cite{Arkani-Hamed:2024jbp} that the cosmohedra are non-simple as polytopes.

In the acyclonesto-cosmohedron, a vertex is a maximal nesting $\tau$ with a maximal nesting $\mathcal{N}$ of its Hasse diagram. Edges incident on that vertex are labelled via dropping a proper nest from $\mathcal{N}$ or by collapsing a minimal nest of $\mathcal{N}$.
In a $d$-dimensional acyclonestohedron, a maximal nesting will have $d+1$ nests. Then the tree (or forest) of these tubes contains at most $\lfloor (d+1)/2 \rfloor$ minimal nests, all a single edge. This maximum occurs for instance when the Hasse diagram is linear, a totally ordered poset. (That in turn does occur if the building set is from a simple graph; it may not be the case when the building set is from a general hypergraph.) \footnote{Thanks to Andrew Sack for pointing out that the Galashin poset associahedra from \cite{galashin} do always contain totally ordered nestings, when there is a single connected component. The poset associahedra from \cite{dfrs} do not always contain totally ordered tubings.} Thus for the acyclonesto-cosmohedron in this case, there is a maximum of $\lfloor (3d-1)/2 \rfloor$ edges incident to such a vertex.
Thus in this case it is always a non-simple polytope  for dimension $d>2.$ 

The minimum degree of a vertex is of course the dimension $d$. That is always seen to occur for some vertices of the acyclonesto-cosmohedron, since we can find vertices $(\tau,\mathcal{N})$ where $\mathcal{N}$ is totally nested (for each tree). In this case there is only one minimal nest, and so the number of incident edges is $d-1+1 = d.$

\subsection{ABHY-like realisation of acyclonesto-cosmohedra and cosmological amplitubes}
Acyclonesto-cosmohedra may be realised in an ABHY-like fashion as intersections of half-spaces similar to acyclonestohedra themselves; this construction generalises that for graph cosmohedra given in \cite{Glew:2025otn}. 
Suppose that we are given an oriented building set \((S,\mathcal B,\mathcal C)\) where \((S,\mathcal C)\) is realised by a collection of vectors \(a_i\in V^*\) that span a finite-dimensional real vector space \(V^*\).
For each acyclic nesting \(\tau\subseteq\mathcal B\), define the kinematic variable \(Y_\tau\colon V\to\mathbb R\) as the affine function
\begin{equation}
    Y_\tau = \sum_{B\in\tau}X_B - \sum_{B\in\tau} \delta_{B\setminus\bigcup\{N\in\tau\,|\,N\subsetneq B\}},
\end{equation}
where \(\delta_{B\setminus\bigcup\{N\in\tau\,|\,N\subsetneq B\}}\) is a positive real number (additional cut parameters) associated to the subset \(B\setminus\bigcup\{B'\in\tau\,|\,B'\subsetneq B\}\), and where \(X_B\) was defined in \eqref{eq:acyclonestohedron_function}.

Then the ABHY-like realisation of the acyclonesto-cosmohedron is given by the set of points \(v\in V\) such that
\begin{equation}
\begin{aligned}
    Y_\tau(v)&\ge0\text{ for every acyclic nesting \(\tau\subseteq\mathcal B\)}\\
    X_\kappa(v)&=0\text{ for every \(\kappa\in\max\mathcal B\)}.
\end{aligned}
\end{equation}
(Of course, one also always has the additional equations \eqref{eq:linear_dependence_amongst_X} for the linear dependence amongst the \(X_B\).)
For this to realise the acyclonesto-cosmohedron, there are additional inequalities that must be satisfied by the cut parameters \(\delta\) and \(c\); it suffices to have 
\begin{align}\label{eq:cosmohedron_inequalities}
    \delta_{S'} &\ll \delta_{S''},&
    \delta_{S'}&\ll c_B
\end{align}
whenever \(|S'|<|S''|\) and for arbitrary \(B\), in addition to \eqref{eq:sufficient_condition_acyclonestohedron}.

In the next section we calculate convincing examples, but finding a general proof that our construction gives the same face poset as the combinatorial definition should not be difficult.

\textbf{Proof sketch}: The proof for the validity of our realization is straightforward for the case of poset cosmohedra, starting with the realization of the graph cosmohedra as defined in \cite{Glew:2025otn} and following the logic of the proof for the poset associahedra in \cite{mppfull}. Note that the cosmohedron for a cycle-free, tree-like, poset is precisely the graph cosmohedron for the line graph of the Hasse diagram of that poset. For the acyclic restriction of a general poset we show 1) that the nested nestings of the poset cosmohedron are combinatorially equivalent as a poset to the acyclic restriction of the associated line graph cosmohedron; 2) that our realization is actually a cross section of the graph cosmohedron (an intersection with the hyperplane determined by the cycle equalities), which we also state below as a separate conjecture;  and moreover 3) that this cross section intersects faces of the graph cosmohedron if and only if those cells correspond to acyclic nestings.

Similarly to amplitubes for acyclonestohedra, to the acyclonesto-cosmohedra we may associate rational functions called the \emph{cosmological amplitubes}:
\begin{equation}
    \Psi_{(S,\mathcal B,\mathcal C)}
    =\prod_{(\tau,\mathcal N)}\prod_{N\in\mathcal N}\frac1{\mathcal R_N},
\end{equation}
where \(\mathcal R_N\) are new formal variables associated to each \(N\subseteq\mathcal B\times\mathcal B\). (Recall from section~\ref{ssec:cosmohedron_definition} that these may be identified with regions in the sense of \cite{Arkani-Hamed:2024jbp,Glew:2025otn}.)
These may be read off from the canonical form of the acyclonesto-cosmohedron in the same manner as for graph cosmohedra \cite[§3.5]{Glew:2025otn}.
\begin{figure}
\centering
    \includegraphics[scale=0.38]{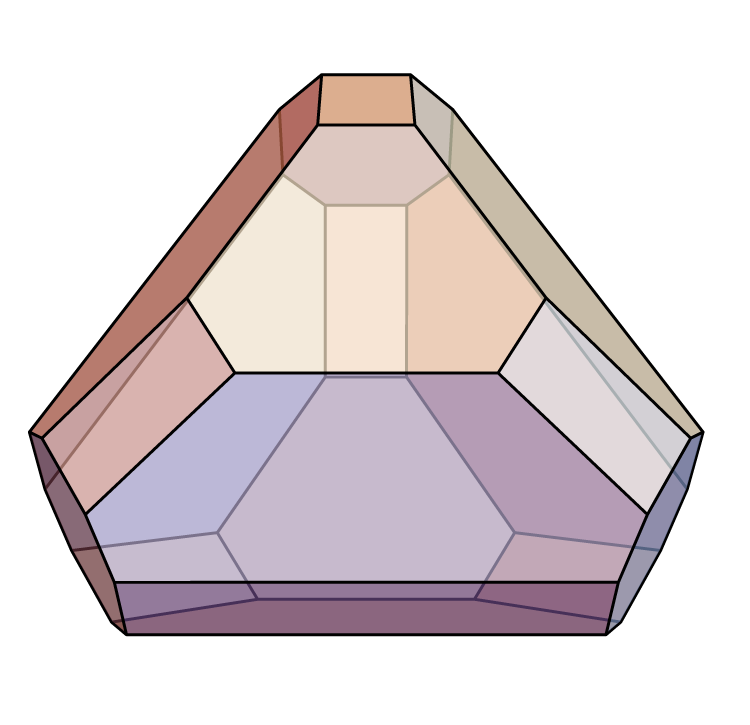} \quad \quad \quad \quad
    \includegraphics[scale=0.415]{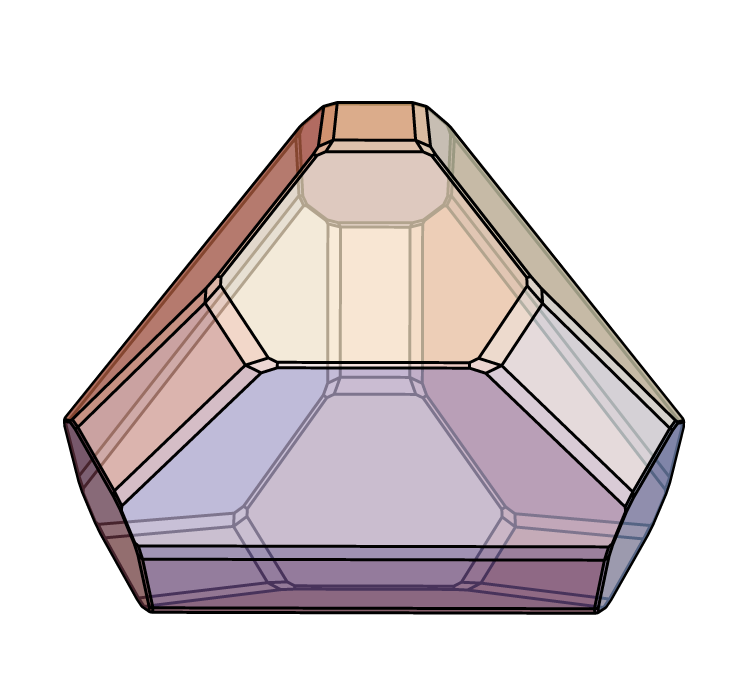}
    \caption{Explicit realisations for the acyclonestohedron and its associated acyclonesto-cosmohedron for the $K_{2,3}$ poset discussed in Example~\ref{ex:K23}.}\label{fig:RealK23}
\end{figure}
\subsection{Examples of acyclonesto-cosmohedra}
As the combinatorial explosion becomes extreme for the acyclonesto-cosmohedron, we explain in detail only the acyclonesto-cosmohedron for the diamond poset; in addition, a portion of the cosmohedron for the Stasheff associahedron together with a diagrammatic depiction of the nested nestings is given in figure~\ref{fig:stasheff_cosmohedron}, and the acyclonesto-cosmohedron for the permutahedron (which is the permutoassociahedron) is shown in figure~\ref{fig:permutahedron_cosmohedron}.
\begin{example}[Diamond poset]
    Consider the diamond poset
    \begin{equation}P_\diamond=
    \begin{tikzcd}[column sep=0,row sep=1em]
    &\bullet\dlar[dash,"\mathtt a"']\drar[dash,"\mathtt b"]\\\bullet\drar[dash,"\mathtt c"']&&\bullet\dlar[dash,"\mathtt d"]\\
    &\bullet
    \end{tikzcd}\end{equation}
    from example~\ref{ex:diamond}, whose oriented building set is
    \begin{align}
        S &= \{\mathtt a,\mathtt b,\mathtt c,\mathtt d\},&
        \mathcal C &= \left\{
            \pm\mathtt a
            \pm\mathtt c\mp\mathtt d\mp\mathtt b
        \right\},&
        \mathcal B &= \mathcal P(S) \setminus \left\{\varnothing,\{\mathtt a,\mathtt d\},\{\mathtt b,\mathtt c\}\right\},
    \end{align}
    with 13 nestings:
    \begin{itemize}
        \item one trivial nesting: \(\{S\}\)
        \item six nestings that correspond to facets of the acyclonestohedron, of the form \(\{S,B\}\) for \(B\in\mathcal B\setminus \left\{\{\mathtt a,\mathtt c\},\{\mathtt b,\mathtt d\}\right\}\) with \(|B|\le2\). (The would-be facets \(\{S,\{\mathtt a,\mathtt c\}\}\) and \(\{S,\{\mathtt b,\mathtt d\}\}\) violate the acyclicity condition.)
        \item six nestings that correspond to vertices of the acyclonestohedron:
        \[
        \begin{gathered}
            \{S,\{\mathtt a,\mathtt b\},\{\mathtt a\}\},\qquad\{S,\{\mathtt a,\mathtt b\},\{\mathtt b\}\},\qquad\{S,\{\mathtt c,\mathtt d\},\{\mathtt c\}\},\\\{S,\{\mathtt c,\mathtt d\},\{\mathtt d\}\},\qquad\{S,\{\mathtt a\},\{\mathtt d\}\},\qquad\{S,\{\mathtt b\},\{\mathtt c\}\}.
        \end{gathered}
        \]
    \end{itemize}
    The Hasse diagrams for all nestings are path graphs, so that we have the following nested nestings:
    \begin{itemize}
        \item one trivial nested nesting: \((\{S\},\varnothing)\)
        \item the six nestings that correspond to facets of the acyclonestohedron each admit a unique nested nesting. For example, to the nesting \(\{S,\{\mathtt a,\mathtt b\}\}\), we can associate the nested nesting \((\{S,\{\mathtt a,\mathtt b\}\},\{\{(\{\mathtt a,\mathtt b\},S)\}\})\).
        \item the six nestings that correspond to facets of the acyclonestohedron each correspond to threes nested nestings, for 18 total. For example, the nesting \(\{S,\{\mathtt a\},\{\mathtt d\}\}\) corresponds to the nested nestings \[(\{S,\{\mathtt a\},\{\mathtt d\}\},\{\{(\{\mathtt a\},S)\},\{(\{\mathtt a\},S),(\{\mathtt d\},S)\}\})\] and \[(\{S,\{\mathtt a\},\{\mathtt d\}\},\{\{(\{\mathtt d\},S)\},\{(\{\mathtt a\},S),(\{\mathtt d\},S)\}\})\] and \[(\{S,\{\mathtt a\},\{\mathtt d\}\},\{\{(\{\mathtt a\},S),(\{\mathtt d\},S)\}\}).\]
    \end{itemize}
    \begin{figure}
\centering
    \includegraphics[scale=0.43]{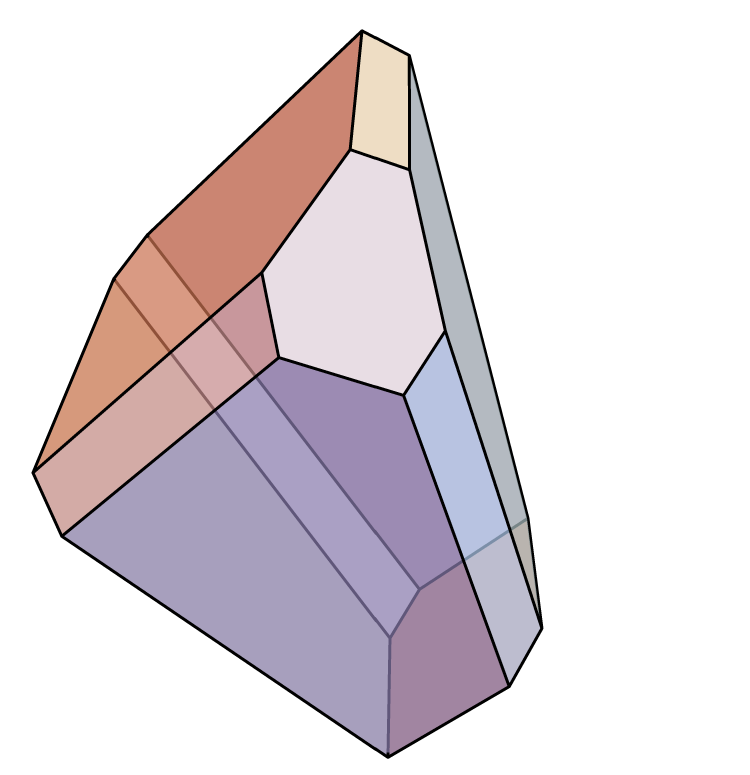} \quad \quad 
    \includegraphics[scale=0.425]{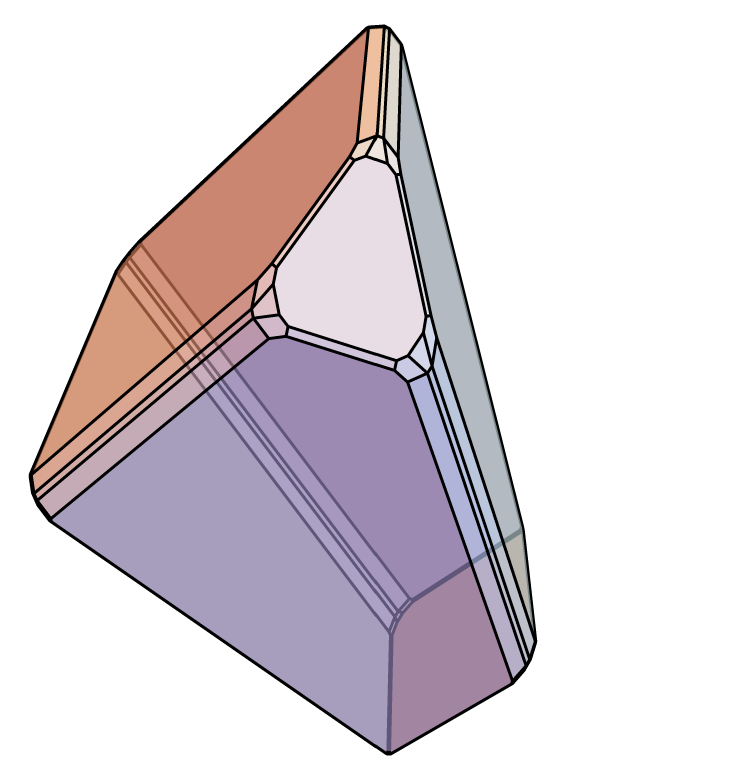}
    \caption{The poset associahedron (left) and poset cosmohedron (right) for the poset depicted in Figure~$1$ of \cite{galashin}. Both polytopes are obtained as sections of the graph associahedron/cosmohedron for the line graph of the Hasse diagram.}
    \label{fig:galash}
\end{figure}
    Therefore, the acyclonesto-cosmohedron for the diamond poset is a dodecagon.
    The ABHY-like realisation of the corresponding acyclonesto-cosmohedron is as follows:
    \begin{equation}
    \begin{aligned}
        Y_{\{S,\mathtt{\{a,b\},\{a\}}\}}=X_S+2X_{\{\mathtt a\}}+X_{\{\mathtt b\}}-c_{{\{\mathtt{a,b}\}}}-\delta_{\{\mathtt a\}}-\delta_{\{\mathtt b\}}-\delta_{\{\mathtt{c,d}\}}&\ge 0\\
        Y_{\{S,\mathtt{\{a,b\},\{b\}}\}}=X_S+X_{\{\mathtt a\}}+2X_{\{\mathtt b\}}- c_{\{\mathtt{a,b}\}}-\delta_{\{\mathtt a\}}-\delta_{\{\mathtt b\}}-\delta_{\{\mathtt{c,d}\}}&\ge0\\
        Y_{\{S,\{\mathtt{a,b}\}\}}=X_S+X_{\{\mathtt a\}}+X_{\{\mathtt b\}}-c_{\{\mathtt{a,b}\}}-\delta_{\{\mathtt{a,b}\}}-\delta_{\{\mathtt{c,d}\}}&\ge0\\
        Y_{\{S,\mathtt{\{c,d\},\{c\}}\}}=X_S+2X_{\{\mathtt c\}}+X_{\{\mathtt d\}}-c_{\{\mathtt{c,d}\}}-\delta_{\{\mathtt c\}}-\delta_{\{\mathtt d\}}-\delta_{\{\mathtt{a,b}\}}&\ge 0\\
        Y_{\{S,\mathtt{\{c,d\},\{d\}}\}}=X_S+X_{\{\mathtt c\}}+2X_{\{\mathtt d\}}-c_{\{\mathtt{c,d}\}}-\delta_{\{\mathtt c\}}-\delta_{\{\mathtt d\}}-\delta_{\{\mathtt{a,b}\}}&\ge 0\\
        Y_{\{S,\{\mathtt{c,d}\}\}}=X_S+X_{\{\mathtt c\}}+X_{\{\mathtt d\}}-c_{\{\mathtt{a,b}\}}-\delta_{\{\mathtt{a,b}\}}-\delta_{\{\mathtt{c,d}\}}&\ge0\\
        Y_{\{S,\mathtt{\{a\},\{d\}}\}}=X_S+X_{\{\mathtt a\}}+X_{\{\mathtt d\}}-\delta_{\{\mathtt{b,c}\}}-\delta_{\{\mathtt a\}}-\delta_{\{\mathtt d\}}&\ge0\\
        Y_{\{S,\mathtt{\{b\},\{c\}}\}}=X_S+X_{\{\mathtt b\}}+X_{\{\mathtt c\}}-\delta_{\{\mathtt{a,d}\}}-\delta_{\{\mathtt b\}}-\delta_{\{\mathtt c\}}&\ge0\\
        Y_{\{S,\{\mathtt a\}\}}=X_S+X_{\{\mathtt a\}}-\delta_{\{\mathtt a\}}-\delta_{\{\mathtt{b,c,d}\}}&\ge0\\
        Y_{\{S,\{\mathtt b\}\}}=X_S+X_{\{\mathtt b\}}-\delta_{\{\mathtt b\}}-\delta_{\{\mathtt{a,c,d}\}}&\ge0\\
        Y_{\{S,\{\mathtt c\}\}}=X_S+X_{\{\mathtt c\}}-\delta_{\{\mathtt c\}}-\delta_{\{\mathtt{a,b,d}\}}&\ge0\\
        Y_{\{S,\{\mathtt d\}\}}=X_S+X_{\{\mathtt d\}}-\delta_{\{\mathtt d\}}-\delta_{\{\mathtt{a,b,c}\}}&\ge0\\
        X_S=X_{\{\mathtt a\}}+X_{\{\mathtt b\}}+X_{\{\mathtt c\}}+X_{\{\mathtt d\}}-c_{\{\mathtt{a,b}\}}-c_{\{\mathtt{c,d}\}}-c_{\{\mathtt{a,b,c,d}\}}&=0\\
        X_{\{\mathtt a\}}-X_{\{\mathtt b\}}+X_{\{\mathtt c\}}-X_{\{\mathtt d\}}&=0.
    \end{aligned}
    \end{equation}
    With \(C\) as in \eqref{eq:square_size_for_diamond}, 
    this reduces to
    \begin{equation}
    \begin{gathered}
        \delta_{\{\mathtt a\}}+\delta_{\{\mathtt{b,c,d}\}}\le X_{\mathtt a}\le C-\delta_{\{\mathtt c\}}-\delta_{\{\mathtt{a,b,d}\}}\\
        \delta_{\{\mathtt b\}}+\delta_{\{\mathtt{a,c,d}\}}\le X_{\mathtt b}\le C-\delta_{\{\mathtt d\}}-\delta_{\{\mathtt{a,b,c}\}}\\
        X_{\{\mathtt a\}}+X_{\{\mathtt b\}}\ge c_{\{\mathtt{a,b}\}}+\delta_{\{\mathtt{a,b}\}}+\delta_{\{\mathtt{c,d}\}}\\
        (C-X_{\{\mathtt a\}})+(C-X_{\{\mathtt d\}})\ge c_{\{\mathtt{c,d}\}}+\delta_{\{\mathtt{a,b}\}}+\delta_{\{\mathtt{c,d}\}}\\
        2X_{\{\mathtt a\}}+X_{\{\mathtt b\}}\ge c_{\{\mathtt{a,b}\}}+\delta_{\{\mathtt{c,d}\}}+\delta_{\{\mathtt a\}}+\delta_{\{\mathtt b\}}\\
        X_{\{\mathtt a\}}+2X_{\{\mathtt b\}}\ge c_{\{\mathtt{a,b}\}}+\delta_{\{\mathtt{c,d}\}}+\delta_{\{\mathtt a\}}+\delta_{\{\mathtt b\}}\\
        2(C-X_{\{\mathtt a\}})+(C-X_{\{\mathtt b\}})\ge c_{\{\mathtt{c,d}\}}+\delta_{\{\mathtt{a,b}\}}+\delta_{\{\mathtt c\}}+\delta_{\{\mathtt d\}}\\
        (C-X_{\{\mathtt a\}})+2(C-X_{\{\mathtt b\}})\ge c_{\{\mathtt{c,d}\}}+\delta_{\{\mathtt{a,b}\}}+\delta_{\{\mathtt c\}}+\delta_{\{\mathtt d\}}\\
        X_{\{\mathtt a\}}+X_{\{\mathtt d\}}\ge\delta_{\{\mathtt{b,c}\}}+\delta_{\{\mathtt a\}}+\delta_{\{\mathtt d\}},\qquad
        X_{\{\mathtt b\}}+X_{\{\mathtt c\}}\ge\delta_{\{\mathtt{a,d}\}}+\delta_{\{\mathtt b\}}+\delta_{\{\mathtt c\}}        
        ,
    \end{gathered}
    \end{equation}
    which indeed defines a dodecagon provided that the inequalities \eqref{eq:sufficient_condition_acyclonestohedron} and \eqref{eq:cosmohedron_inequalities} hold.
    The corresponding cosmological amplitube has 12 terms corresponding to the 12 vertices of the dodecagon, which we omit.
\end{example}

\begin{example}[Simplex]
    Recall the single edge hypergraph from Example~\ref{ex:simplex}. The Hasse diagrams of the maximal nestings $\tau$ are claw graphs, and nestings $\mathcal{N}$ of those will always be totally nested.  Thus the acyclonesto-cosmohedron in this case will be simple, and will recapture the combinatorics of the permutohedron, as pointed out in  \cite{Glew:2025otn}.
\end{example}

\begin{example}[Bowtie poset]
    Recall the bowtie poset from Example~\ref{ex:bowtie}. Figure\ref{fig:bigbow} shows the combinatorics of the acyclonesto-cosmohedron for that poset, which is a 16-sided polygon. 
\end{example}

\begin{figure}
\centering
    \includegraphics[scale=0.38]{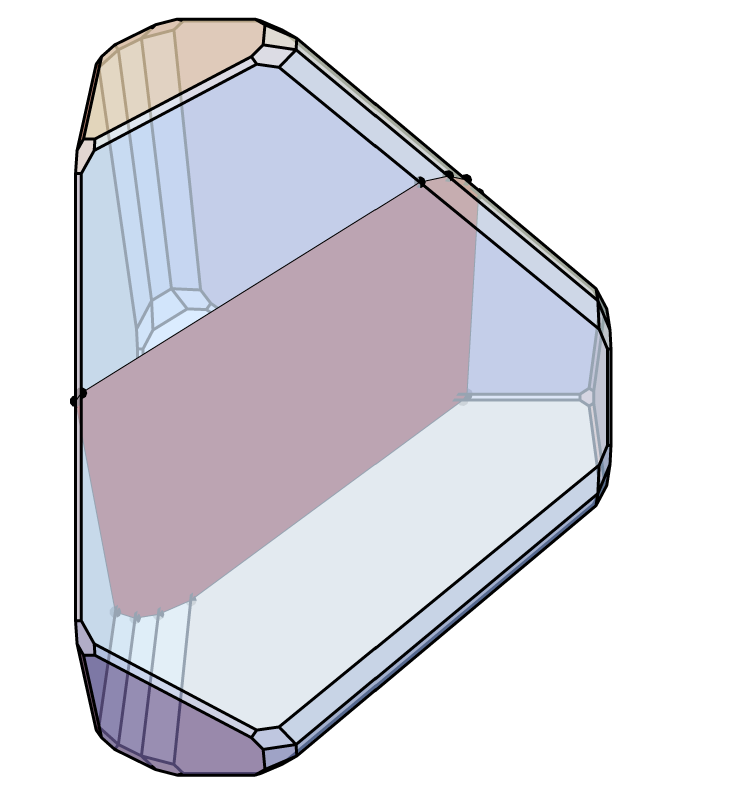} \quad \quad \quad \quad
    \includegraphics[scale=0.38]{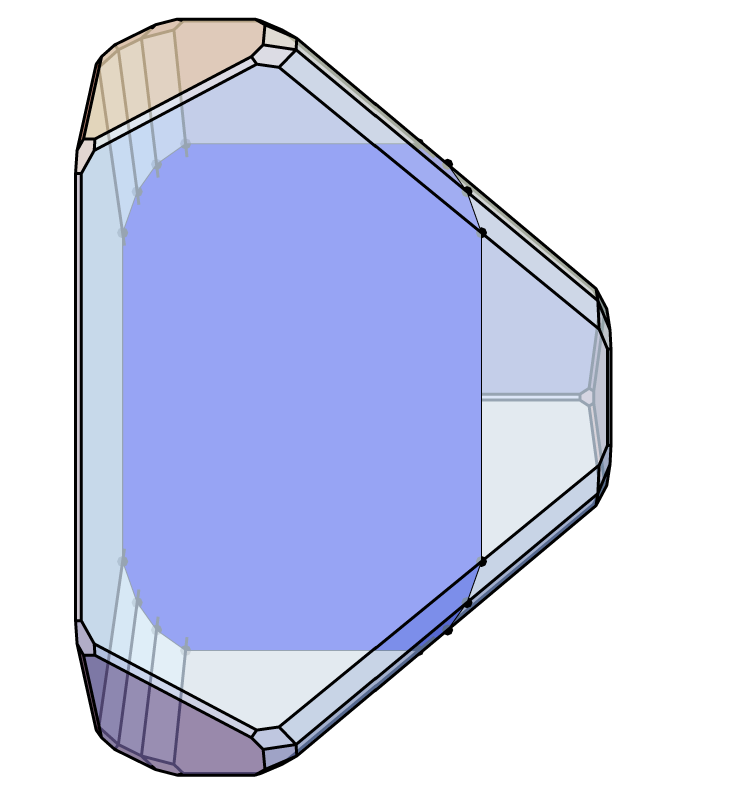}
    \caption{The poset cosmohedron for the diamond poset (left) and the bow-tie poset (right) realised as sections of the graph cosmohedron for the four-cycle.}
    \label{fig:sections}
\end{figure}

\begin{example}[Stasheff associahedron]
Recall the path graph, as a totally ordered set from Example~\ref{ex:stasheff}. The cosmohedron in two dimensions is a decagon, and the three-dimensional cosmohedron is shown in Figure~\ref{fig:stasheff_cosmohedron}.
\end{example}

\begin{example}[Permutohedron]
Recall the claw poset from Example~\ref{ex:permutohedron}. The acyclonesto-cosmohedron in two dimensions is a dodecagon, and the three-dimensional acyclonesto-cosmohedron is shown in Figure~\ref{fig:permutahedron_cosmohedron}. Note that every maximal nesting of the claw poset is totally nested, so that the corresponding facets of the acyclonesto-cosmohedron are copies of the  associahedron. Thus the claw poset acyclonesto-cosmohedra (and its line graph, the complete graph) recaptures the combinatorics of the permutoassociahedron as shown in \cite{Glew:2025otn}.
\end{example}

\begin{example}[$K_{2,3}$]
\label{ex:K23}
    The poset with three maximal nodes, two minimal nodes, and all covering relations between them is pictured in Figure~\ref{fig:k23}. Explicit realisations of both the acyclonestohedron and acyclonesto-cosmohedron for this case are shown in Figure~\ref{fig:RealK23}. Note that the acyclonestohedron has three octagonal facets and its acyclonesto-cosmohedron has three 16-gons. As well, every maximal nesting of the poset $K_{2,3}$ is totally nested, so that the corresponding facets of the acyclonesto-cosmohedron are pentagonal: copies of the two-dimensional associahedron, which is the operahedron on the linear tree. 
\end{example}

\begin{example}[Poset cosmohedra as sections]
    It was shown in \cite{mppfull} that acyclonestohedra can be obtained as sections of graph associahedra. In this example we provide evidence that the same holds true for acyclonesto-cosmohedra, that is, they can be obtained as sections of the graph cosmohedra associated to the line graph of the Hasse diagram. Consider the diamond and bow-tie posets. The line graph of the Hasse diagram for both posets produces the graph $C_4$. This suggests that both poset cosmohedra can be obtained as sections of the same polytope, the graph cosmohedron for $C_4$. We find this is indeed the case, as illustrated in  Figure~\ref{fig:sections}. This leads us to conjecture that all poset cosmohedra can be obtained as sections of graph cosmohedra. Further evidence to support this conjecture is displayed in Figure~\ref{fig:galash}.
\end{example}
\section{Outlook}
As this paper was being finished, the paper \cite{Figueiredo:2025daa} appeared introducing the correlatron, a polytope that interpolates between the cosmohedron and the associahedron. It would be interesting to see if the construction of correlation polytopes extends nicely to graph associahedra and acyclonestohedra.

We also note that our constructions should extend easily to other combinatorial polytopes that are based on a ``nested set'' paradigm. Whenever the faces of the polytope are labelled by certain subsets of a power set, the natural partial ordering on that subset, by inclusion, allows a poset associahedron construction for each face. Thus a cosmohedron construction is available by blowing up each original face to codimension one according to its poset associahedron. One candidate for this process that deserves study is the poset associahedra of \cite{dfrs}, or acyclic versions of those. These poset associahedra contain and generalise nestohedra and graph associahedra, so therefore with the addition of an oriented matroid they generalise the acyclonestohedra as well.

\begin{landscape}
\begin{figure}\label{fig:bowtie}
    \centering
    \includegraphics[width=19.33cm]{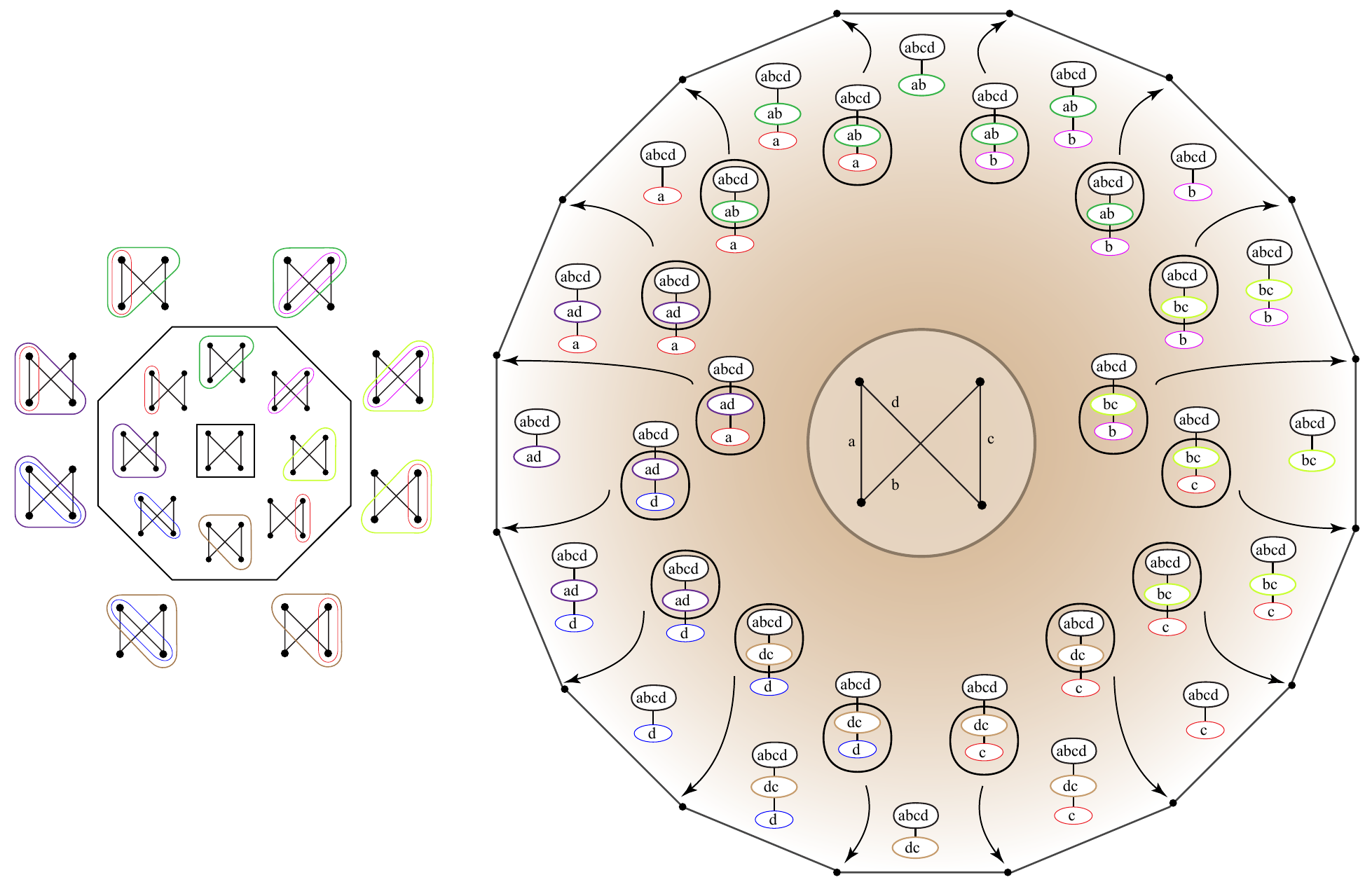}
    \caption{The poset associahedron for the poset in Example~\ref{ex:bowtie} is pictured on the left, and its acyclonesto-cosmohedron is the 16-gon.}
    \label{fig:bigbow}
\end{figure}
\end{landscape}

\begin{landscape}
\begin{figure}
\centering
\includegraphics[width=19.33cm]{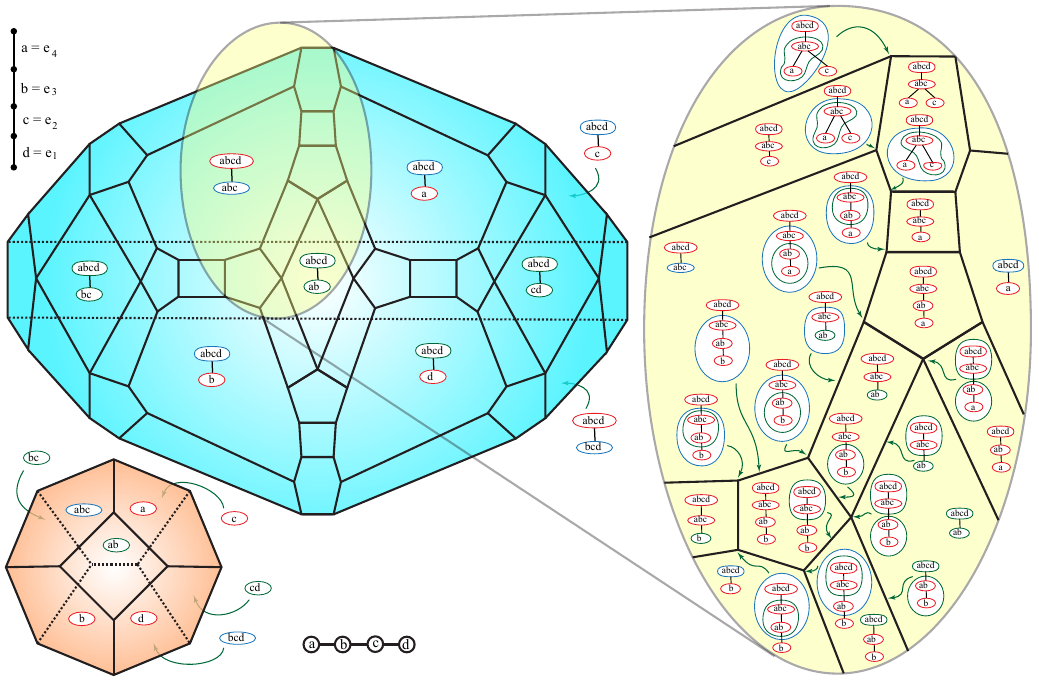}
\caption{Three-dimensional cosmohedron for the Stasheff associahedron. The associahedron (lower left) can be labelled by tubings on a path graph with nodes $\mathtt{a,b,c,d}$, or nestings on the totally ordered poset with edges $\mathtt{a,b,c,d}.$ Every nesting $\tau$ contains the improper nest $\{\mathtt{a,b,c, d}\},$ and every nesting $\mathcal{N}$ on $\tau$ contains the improper nest $\{\tau\}$, although the pictures leave this out. Single proper nests label nine of the facets in the cosmohedron as shown here. Note for comparison that part of the zoomed-in portion here matches part of Figure 8 in \cite{Arkani-Hamed:2024jbp}. } \label{fig:stasheff_cosmohedron}
\end{figure}
\end{landscape}
\begin{landscape}
\begin{figure}
\centering
\includegraphics[width=19.33cm]{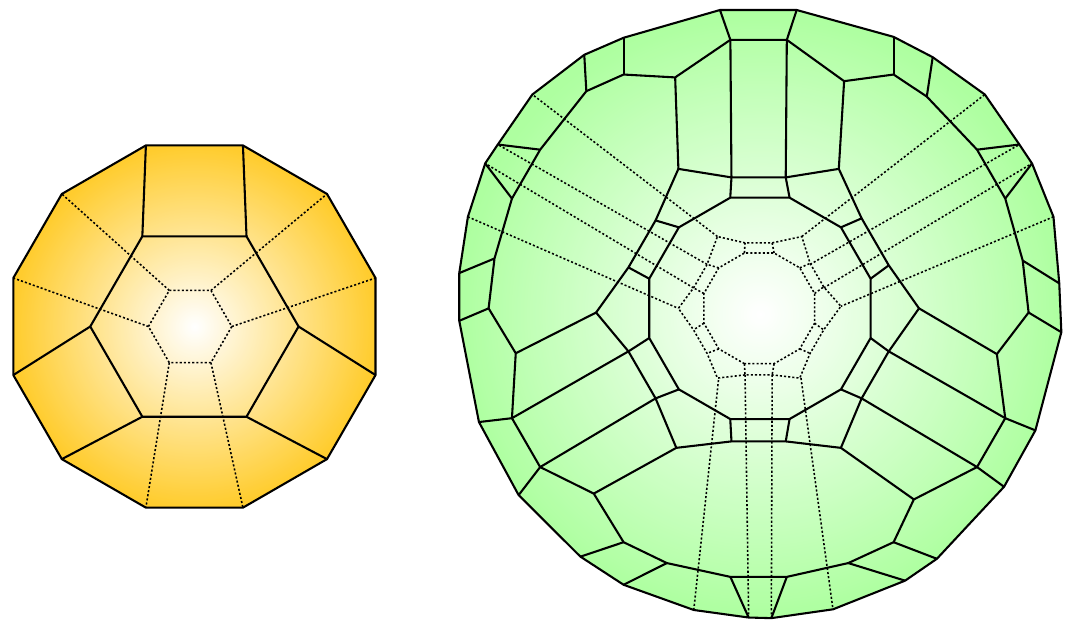}
\caption{Acyclonesto-cosmohedron for the permutahedron: this gives a different view of the permutoassociahedron.}\label{fig:permutahedron_cosmohedron}
\end{figure}
\end{landscape}
\section*{Acknowledgements}
R.G. and H.K. thank Mathison Francis Knight\textsuperscript{\orcidlink{0009-0005-7014-3255}} and Tomasz Łukowski\textsuperscript{\orcidlink{0000-0002-4159-3573}} for helpful discussion. R.G. would also like to thank Andrzej Pokraka\textsuperscript{\orcidlink{0000-0003-1186-4624}}, Nima Arkani-Hamed, and Carolina Figueiredo for discussions on related topics.

\bibliographystyle{alphaurl}
\bibliography{biblio}

\end{document}